\documentclass[aip,pop,reprint,10pt,superscriptaddress,showpacs]{revtex4-1}
\usepackage{amsfonts}
\usepackage{amsmath}
\usepackage{amssymb}
\usepackage{graphicx}
\usepackage{subfigure}
\usepackage{color}
\begin{document}
\title{Oblique propagation of dust ion-acoustic solitary waves in a magnetized dusty pair-ion plasma}
\author{A. P. Misra}
\email{apmisra@visva-bharati.ac.in; apmisra@gmail.com}
\author{Arnab Barman}
\affiliation{Department of Mathematics, Siksha Bhavana, Visva-Bharati University, Santiniketan-731 235, West Bengal, India}
\begin{abstract}
We investigate the  propagation characteristics of electrostatic waves   in a magnetized pair-ion plasma with immobile  charged dusts. It is shown that obliquely propagating (OP) low-frequency (in comparison with the negative-ion cyclotron frequency) long-wavelength ``slow" and ``fast"   modes can propagate, respectively, as  dust ion-acoustic (DIA)  and   dust ion-cyclotron (DIC)-like  waves. The properties of these    modes   are studied with the effects of obliqueness of propagation $(\theta)$, the static magnetic field,  the ratios of the negative to positive ion masses  $(m)$ and   temperatures $(T)$  as well as the     dust to negative-ion number density ratio  $(\delta)$. Using  the standard reductive perturbation technique, we derive    a Korteweg-de Vries (KdV) equation which governs the evolution of small-amplitude OP DIA waves. It is found that the KdV equation   admits  only rarefactive solitons in plasmas with  $m$ well below its critical value $m_c~(\gg1)$ which   typically depends on $T$ and $\delta$. It is shown that  the nonlinear coefficient of the KdV equation vanishes at $m=m_c$, i.e., for plasmas with much heavier negative ions,  and the evolution of the DIA waves   is then described by a modified KdV (mKdV) equation. The latter is  shown  to have only compressive soliton solution.  The  properties of both the KdV and mKdV  solitons are studied with the system parameters as above, and possible applications of our results to laboratory and space plasmas are  briefly discussed.
\end{abstract}
  \date{19 June 2014}
\maketitle
\section{Introduction}
Typical plasmas consisting of electrons and ions or similar particles with large-mass difference essentially cause temporal as well as spatial variations of collective plasma phenomena. However, the space-time parity can be maintained in pair-ion plasmas with  equal mass or a slightly different masses. Plasmas containing positive and negative ions not only found in naturally occurring plasmas, but   are also used  for technological applications. In  many industries, e.g., in integrated-circuit fabrication, since the deposited film is strongly damaged by a  high-energy electron, a plasma source having no energetic electrons  is required. For this purpose, a radio-frequency  plasma source has also been developed \cite{yabe1994}.   Pure pair-ion plasmas with equal mass and temperature have been generated in the laboratory, and   three kinds of electrostatic modes, namely the ion-acoustic wave (IAW), the intermediate-frequency wave (IFW), and the ion plasma wave (IPW), have been experimentally observed using fullerene as an ion source \cite{pair-ion-experiment}. Later, a criterion for the description of pure pair-ion plasmas has been investigated by Saleem \cite{saleem-pair-ion}. Recently, there has been a growing interest in investigating the properties of electrostatic waves in pair-ion plasmas in linear and nonlinear regimes (See, e.g., \cite{pair-ion1,pair-ion2,pair-ion3,pair-ion4,pair-ion5,pair-ion-charging-theory,pair-ion-charging-experiment,pair-ion-instability}). To mention few, Misra \textit{et al.} \cite{pair-ion1} had investigated the   propagation of dust ion-acoustic (DIA) solitary waves and shocks (SWS) in an unmagnetized dusty negative-ion plasma. They found that the SWS exist   with  negative  potential when dusts are positively charged. Rosenberg and Merlino \cite{pair-ion-instability} studied the ion-acoustic instability in a dusty negative-ion plasma. The theory of dust-acoustic solitons in unmagnetized pair-ion-electron plasmas  has been investigated through the description of Korteweg-de Vries (KdV) equation \cite{pair-ion2}. More recently,    the linear and nonlinear properties of both small- and large-amplitude DIA  solitary waves in an unmagnetized   pair-ion   plasma with immobile charged dusts have been studied \cite{pair-ion5}. It has been shown  that pair-ion plasmas with positively charged dusts support   DIA  solitons   only of the rarefactive type.  There are, however, a number of recent works  dealing with the nonlinear properties of solitary waves \cite{solitary-sultana,williams2013,sultana2010,mamun2002,kourakis2004a,kourakis2004b,paul2013,mamun1998,saini2008,bandyopadhyay1999},  electrostatic ion-cyclotron waves \cite{ion-cyclotron-shukla} shocks \cite{shocks1} in other plasma environments.

 The presence of negative ions in the Earth's D and lower E regions at altitudes about $73-90$ km \cite{narcisi1971}  as well as  in Titan's atmosphere \cite{coates2007} have been detected. These negative ions act as a precursors for the formation of massive charged dust particles, which, e.g., in a processing reactor can contaminate the product \cite{choi1993}.   The   in situ measurements of charged particles in the polar mesosphere under nighttime conditions revealed the existence of positively charged dusts which are dominated by both positive and negative ions, and few percentage of electrons \cite{in-situ-Rapp}. Also, it has been shown that such positively charged dust particles are due to the presence of sufficiently heavy and numerous negative ions (i.e.,  $m_n>300$ amu and   $n_{n0}\gtrsim 50n_{e0}$, where $m_n$ is the negative-ion mass, and $n_{e0}$, $n_{n0}$ are, respectively, the number densities of electrons and negative ions) \cite{in-situ-Rapp}. Furthermore, it has been observed that the dust particles injected into a pair-ion plasma (e.g., $K^{+}/SF_6^{-}$ plasmas) can become positively charged when $n_{n0}\gtrsim 500n_{e0}$ \cite{pair-ion-charging-experiment,pair-ion-charging-theory}. On the other hand, there has been a number of observations to detect solitary structures in space plasma environments. To mention few,   the Viking spacecraft and Freja satellite have indicated the presence of electrostatic solitary structures in the Earth's magnetosphere \cite{dovner1994}.  Williams {\it et al.} \cite{williams2006}  have reported the observations of   solitary waves by the Cassini spacecraft in the vicinity of Saturn’s magnetosphere (with ambient magnetic fields $\sim0.1-8000$ nT).  It has also been suggested that the turbulence in Earth’s magnetosheath may account for the observations of many solitary pulses observed there \cite{pickett2004}. It is thus of great interest and importance to extend the theory of ion waves \cite{pair-ion-experiment,pair-ion-charging-experiment}  in magnetized pair-ion plasmas with a background of positively charged dusts.  

 In this wok, we  consider a dusty pair-ion plasma  in presence of an external magnetic field, and use a three-dimensional fluid model for the propagation of DIA waves obliquely to the magnetic field. We show that   apart from the usual DIA and dust ion-cyclotron (DIC) modes, which appear for wave propagation   parallel and perpendicular to the magnetic field, there   also exist obliquely propagating very low-frequency (compared to the negative-ion cyclotron frequency) long-wavelength fast and slow modes,   which are similar to DIC and DIA waves.  Using the standard reductive perturbation technique, we derive KdV and modified KdV (mKdV) equations to describe the evolution of nonlinear DIA waves   in plasmas with $m<m_c$ and  $m\sim m_c$ respectively, where $m$ is the mass ratio between negative and positive ions and $m_c$ is its critical value.  It is shown that the KdV and mKdV equations admit   rarefactive and compressive solitons in plasmas with $m>1$ and $m\gg1$   respectively.
  
\section{Basic Equations}
 We consider the   propagation of electrostatic waves in a magnetized collisionless dusty  plasma  consisting of singly charged  adiabatic positive and negative ions, and positively charged dusts. The latter are  assumed to be of uniform size and immobile, since
we are concerned with the occurrence of DIA and DIC waves, whose phase speeds are much larger than the ion- and dust-thermal speeds,  on a time scale much shorter than the
dust plasma and dust gyroperiods. Thus, the charged dust grains do not have time to respond to the  DIA and DIC oscillations, and subsequently there are insignificant dust number density perturbations. However, the effects of the positively charged  dusts  then appear
 through the modification of the charge neutrality condition: 
\begin{equation}
n_{p0}+z_{d}n_{d0}=n_{n0}, \label{charge-neutrality1}
\end{equation}
where $n_{j0}$ is the unperturbed number density of species $j$ ($j$=$p$, $n$, $d$, respectively, stand for  positive ions, negative ions, and static charged dusts), $z_d$ $(>0)$ is the unperturbed dust charge state.  

We consider a hydrodynamic model in which  the negative ion fluids are  heavier than the positive ions, and negative ion number density is much larger than that of electrons so that  dusts become positively charged \cite{pair-ion-charging-experiment,pair-ion1,pair-ion5}.   Thus, the dominant higher mobility species in the plasma are  the positive ions. In DIA and DIC waves, since the thermal motion of ions can not keep up with the wave, both positive and negative ions are adiabatically compressed, and we assume  the adiabatic compression of the ion fluids \cite{pair-ion1,pair-ion5}.   Furthermore,  since the negative ion number density is larger than that of the positive ions [Eq. \eqref{charge-neutrality1}] in presence of positively charged dusts, the phase speed of the DIA wave is somewhat enhanced in comparison with that of the ion-acoustic waves in pair-ion plasmas without charged dusts.  The wave propagation is considered  in an arbitrary direction with respect to the static  magnetic filed $\mathbf{B}=B_0\hat{z}$, and ions are magnetized.  
 
The basic equations for the dynamics of positive and negative ions in presence of the static magnetic field  are
\begin{equation}
\frac{\partial n_j}{\partial t}+\nabla\cdot(n_j\mathbf{v}_j)=0,\label{cont-eqn}
\end{equation}
\begin{equation}
\frac{d\mathbf{v}_j}{dt}=\frac{q_j}{m_j}\left(\mathbf{E}+\mathbf{v}_j\times\frac 1cB_0 \hat z\right)-\frac{\nabla P_j}{m_j n_j},\label{moment-eqn}
\end{equation}
\begin{equation}
\nabla\cdot\mathbf{E}=4\pi e(n_p-n_n+z_d n_{d0}),\label{poisson-eqn}
\end{equation}
where $d\mathbf{v}_j/dt\equiv\partial/\partial t+\mathbf{v}_j\cdot\nabla$, and $n_j$, $\mathbf{v}_j$, and $m_j$, respectively, denote the number density, velocity, and mass of $j$-species particles [$j=p(n)$ stands for positive (negative) ions]. Also, $q_p=e$, $q_n=-e$, $e(>0)$ being the elementary charge, and $\mathbf{E}=-\nabla\phi$ with $\phi$ denoting the electrostatic potential. 

Equations \eqref{cont-eqn} to \eqref{poisson-eqn} can be recast in terms of dimensionless variables. Thus, we normalize the physical quantities according to
$\phi\rightarrow e\phi/k_BT_p$, $n_j\rightarrow n_j/n_{j0}$, $v_j\rightarrow v_j/c_s$, where $c_s=\sqrt{k_BT_p/m_n}=\omega_{pn}\lambda_D$ is the ion-acoustic speed with $\omega_{pn}=\sqrt{4\pi n_{n0}e^2/m_n}$ and $\lambda_D=\sqrt{k_BT_p/4\pi n_{n0}e^2}$ denoting, respectively, the negative-ion plasma frequency and the plasma Debye length. Here, $T_j$ is the thermodynamic temperature of j-species ions and $k_B$ is the Boltzmann constant. The space and time variables are normalized by  $\lambda_D$  and  $\omega^{-1}_{pn}$  respectively. In Eq. \eqref{moment-eqn}, we consider the adiabatic pressure law as \cite{pair-ion1,pair-ion5} $P_j/P_{j0}=(n_j/n_{j0})^\gamma$ with $P_{j0}=n_{j0}k_B T_j$ for each  ion-species $(j=p,n)$, and the adiabatic index $\gamma=5/3$ $[=(2+D)/D$, $D$ being the number of degrees of freedom]. It is to be mentioned that in the propagation of very low-frequency waves (in comparison with the negative-ion cyclotron frequency),  the phase speed $v_p$ of the DIA waves is to be much higher than the thermal speed $v_{tn}$ of negative ions as well as much lower than the positive-ion thermal speed $v_{tp}$, i.e., $v_{tn}\ll v_p\ll v_{tp}$, in order to avoid the wave damping due to the resonance with positive or negative ions. This feature can be verified from  the   linear dispersion relation  to be obtained in the next section.    \\ 
Thus, from Eqs. \eqref{cont-eqn} to \eqref{poisson-eqn}, we obtain the following set of equations in dimensionless form: 
\begin{equation}
\frac{\partial n_j}{\partial t}+\nabla\cdot(n_j\mathbf{v}_j)=0,\label{cont-eqn-nond}
\end{equation}
\begin{eqnarray}
\frac{d\mathbf{v}_j}{dt}=&&-\zeta_j\left(\frac{m_n}{m_j}\nabla\phi-\mathbf{v}_j\times\omega _{cj}\hat z\right)\notag\\
&&-\frac53\frac{m_n T_j}{m_jT_p}n^{-{1/3}}_p\nabla n_j,\label{moment-eqn-nond}
\end{eqnarray}
\begin{equation}
\nabla^2\phi=n_n-\mu n_p-\delta, \label{poisson-eqn-nond}
\end{equation}
together with the charge neutrality condition given by
\begin{equation}
\mu+\delta=1, \label{charge-neutrality-nond}
\end{equation}
where $\zeta_p=1$, $\zeta_n=-1$, $\omega_{cj}=eB_0/cm_j\omega_{pn}$ is the cyclotron frequency for the $j$-th species of ions, normalized by the negative-ion plasma frequency,  $\mu=n_{p0}/n_{n0}$, and  $\delta=z_dn_{d0}/n_{n0}$.  
\section{Dispersion relation: Linear waves}
In the linearization process (very small-amplitude limit), we split-up the physical quantities into its unperturbed and perturbed parts according to $n_j=1+n_{j1}$, $\mathbf{v}_j=\mathbf{v}_{j1}$, $\phi=\phi_1$. The perturbations  are then considered  to vary as $\varpropto\exp(i\mathbf{k}\cdot \mathbf{r}-i\omega t)$, i.e., in the form of oscillations with wave frequency  $\omega$ and wave number $k$.  Thus, Fourier analyzing  Eqs. \eqref{cont-eqn-nond}-\eqref{poisson-eqn-nond} we obtain (omitting the subscripts $1$) 
\begin{equation}
-\omega n_j+\mathbf{k}\cdot\mathbf{v}_j=0,\label{cont-linear}
\end{equation}
\begin{equation}
-\omega\mathbf{v}_p+i\left(\mathbf{v}_p\times\omega _{cp}\hat z\right)+m\mathbf{k}\left( \phi+\frac53n_p\right)=0,\label{moment-linear-p}
\end{equation}
\begin{equation}
-\omega\mathbf{v}_n-i\left(\mathbf{v}_n\times\omega _{cn}\hat z\right)+\mathbf{k}\left( -\phi+\frac53Tn_n\right)=0,\label{moment-linear-n}
\end{equation}
\begin{equation}
n_n-\mu n_p+k^2\phi=0, \label{poisson-linear}
\end{equation}
where we denote $m=m_n/m_p$ and $T=T_n/T_p$ as the mass and temperature ratios of negative to positive ions.
Taking dot product of Eq. \eqref{moment-linear-p} with  $\mathbf{k}$, and using Eq. \eqref{cont-linear} we obtain 
\begin{equation}
n_p\left(-\omega^2+\frac53mk^2\right)+mk^2\phi+i\omega_{cp}\left(k_xv_{py}-k_yv_{px}\right)=0,\label{moment-linear-p-cont-linear}
\end{equation}
From the $x$- and $y$-components of  Eq. \eqref{moment-linear-p}, we solve for $v_{px}$ and $v_{py}$ to yield the following expressions
\begin{equation}
v_{px}=m\left[\frac{\omega k_x+i\omega_{cp} k_y}{\omega^2-\omega^2_{cp}}\right]\left(\phi+\frac53n_p\right),\label{velocity-x-p}
\end{equation}
\begin{equation}
v_{py}=m\left[\frac{\omega k_y-i\omega_{cp} k_x}{\omega^2-\omega^2_{cp}}\right]\left(\phi+\frac53n_p\right).\label{velocity-y-p}
\end{equation}
Substituting the expressions from  Eqs. \eqref{velocity-x-p} and \eqref{velocity-y-p} into Eq. \eqref{moment-linear-p-cont-linear}, we obtain the following expression for the perturbed number density of positive ions
\begin{equation}
n_p=\frac{mA_p}{B_p}\phi,\label{Ap-Bp}
\end{equation}
where $A_p$ and $B_p$ are given by
\begin{equation}
A_p=\omega^2k^2-\omega^2_{cp}k^2_z,\label{A_p}
\end{equation}
\begin{equation}
B_p=\omega^2\left(\omega^2-\frac{5}{3}mk^2\right)-\omega^2_{cp}\left(\omega^2-\frac{5}{3}mk_z^2\right).\label{B_p}
\end{equation}
Proceeding in the same way as  for $n_p$ of positive ions, we   obtain from Eqs. \eqref{cont-linear} and \eqref{moment-linear-n} the similar expression for the negative ions as
\begin{equation}
n_n=-\frac{A_n}{B_n}\phi,\label{An-Bn}
\end{equation}
where $A_n$ and $B_n$ are given by
\begin{equation}
A_n=\omega^2k^2-\omega^2_{cn}k^2_z,\label{A_n}
\end{equation}
\begin{equation}
B_n=\omega^2\left(\omega^2-\frac{5}{3}Tk^2\right)-\omega^2_{cn}\left(\omega^2-\frac{5}{3}Tk_z^2\right).\label{B_n}
\end{equation}

Next, substituting the expressions for $n_n$ and  $n_p$ from Eqs. \eqref{Ap-Bp} and \eqref{An-Bn} into Eq. \eqref{poisson-linear}, and assuming that the perturbation $\phi$ is nonzero,    we obtain the following linear dispersion relation:
 \begin{equation}
\sum_{j=p,n}{\frac{\left(m_nn_{j0}/m_jn_{n0}\right) f_{cj}}{\omega^2\left(1-g_{cj}\right)-5k_z^2v^2_{tj}/3}}=1,\label{disp-relation}
\end{equation}
where  $f_{cj}=k^2_z/k^2-\omega^2/\omega^2_{cj}$  and $g_{cj}=\left(\omega^2-{5}k^2v^2_{tj}/3\right)/\omega^2_{cj}$ in which $v_{tj}=\sqrt{k_BT_j/m_j}/c_s$ is the thermal velocity for $j$-species of ions, normalized by $c_s$. Note, however, that the second term of $f_{cj}$, and  $g_{cj}$   appear due to the effects of the static magnetic field and the velocity perturbations transverse to it.    By disregarding these effects, and considering one-dimensional wave propagation (Hence replacing the factor $5/3$ by $3$ as the contribution from the thermal adiabatic pressure in one-dimensional propagation), one  can recover the dispersion relation for dust ion-acoustic waves in unmagnetized plasmas (See Eq. (9) of Ref. \cite{pair-ion5}). The dispersion relation   \eqref{disp-relation} thus generalizes and extends the work of Ref. \cite{pair-ion5}    to  provide some new wave  modes that were not reported before.  From this dispersion relation, we  not only recover the usual  DIA and  DIC    modes, but also some other low-frequency (compared to the negative-ion cyclotron frequency) long-wavelength DIA and DIC-like modes. On the left-side of Eq. \eqref{disp-relation}, the first (second) term is the contribution from positive (negative) ion fluids whose equilibrium are under the electrostatic force, the Lorentz force, and the adiabatic pressure, whereas the nonzero  value $1$ on the right-hand side is from the higher-order dispersive effects   due to separation of charged particles (Deviation from quasi-neutrality in the perturbed state).  From the dispersion relation  \eqref{disp-relation} we  find, in particular, that in order to avoid the wave   damping  for very low-frequency waves ($\omega^2\ll\omega^2_{cn}\ll\omega^2_{cp}$  with  $g_{cj}\ll1$) due to the resonance with either the positive or negative ions, the phase speed of the wave is to be much higher than the negative-ion thermal speed $v_{tn}$ and much lower than the positive-ion thermal speed $v_{tp}$. Such wave damping may arise in an unmagnetized pair-ion plasma \cite{pair-ion5} or plasmas with one group of ions \cite{ppcf-sultana}.  The dispersion equation  \eqref{disp-relation} is of  degree eight in $\omega$, and,  in general, can give eight wave modes.  However, we will be interested to study the properties of  some useful wave modes as mentioned above at some interesting limiting cases as discussed below. 

\textit{First of all}, in the quasineutrality limit (valid for long-wavelength modes) and in the low-frequency approximation, i.e.,  $\omega^2\ll\omega^2_{cn}\ll\omega^2_{cp}$, we obtain from  Eq. \eqref{disp-relation} the following wave modes
\begin{widetext}
\begin{equation}
\omega^2\approx\frac12\left(\frac{m\mu}{\omega^2_{cp}}+\frac{1}{\omega^2_{cn}}\right)^{-1}\left[\alpha_1+\alpha_2k^2\pm\sqrt{\left(\alpha_1-\alpha_2k^2\right)^2+\frac{20}{3}m\mu (m-T)\left(\frac{1}{\omega^2_{cn}}-\frac{1}{\omega^2_{cp}}\right)k^2 \cos^2\theta}\right], \label{low-freq-mode}               
\end{equation}
\end{widetext}
where 
\begin{equation}
\alpha_1=\left(1+m\mu\right)\cos^2\theta,\hskip5pt \alpha_2=\frac53m\left(\frac{\mu T}{\omega^2_{cp}}+\frac{1}{\omega^2_{cn}}\right), \label{alpha1-alpha2}   
\end{equation}
and ${k_z}={k}\cos\theta$, $\theta$ being the angle between the magnetic field and the wave vector. The upper (plus) and lower (minus) signs in Eq. \eqref{low-freq-mode} represent, respectively, the electrostatic low-frequency fast and slow waves which   propagate obliquely  to the external magnetic field.   The phase velocities of these waves approach  a maximum value in the long-wavelength limits.

\begin{figure*}[ht]
\centering
\subfigure[]{\includegraphics[height=2.1in,width=2.26in]{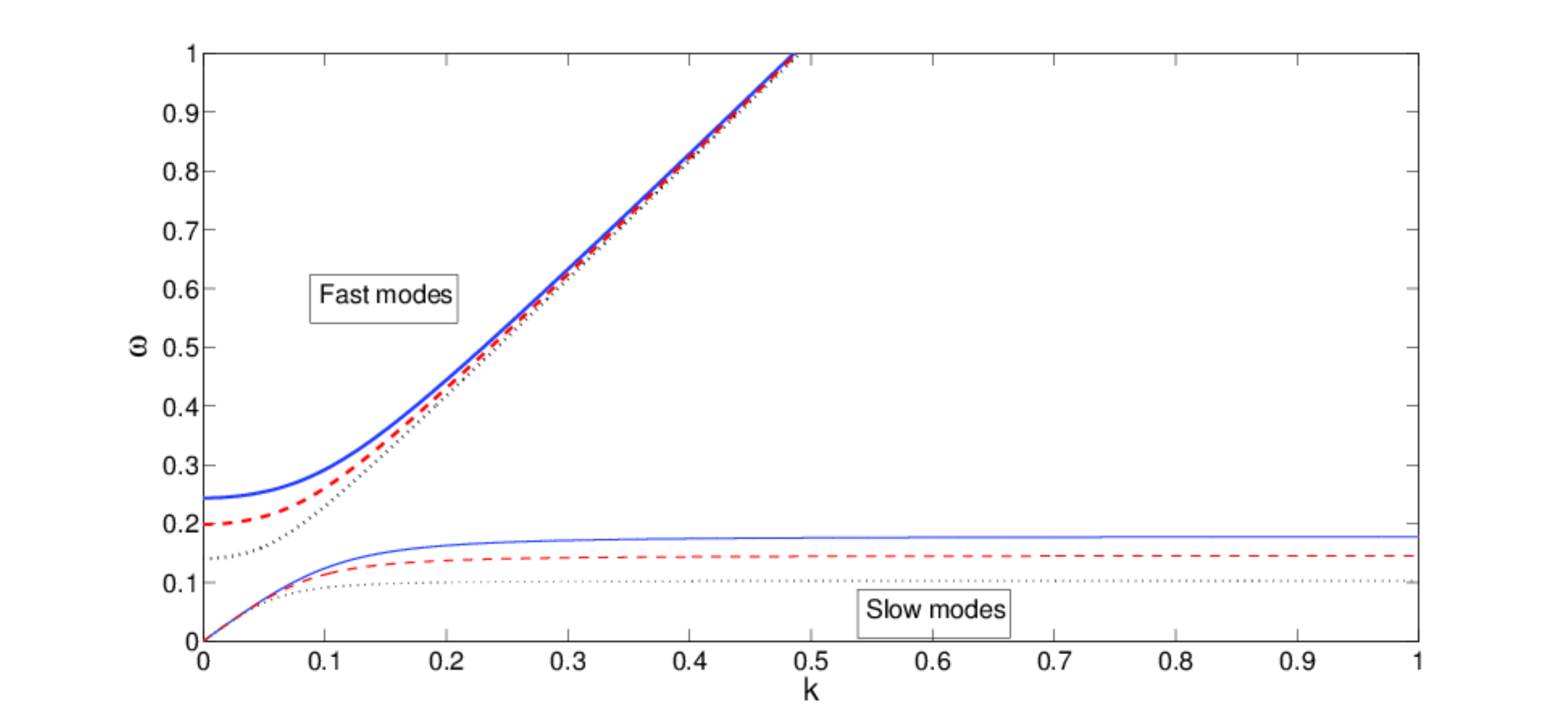}
\label{fig:subfigure1a}}
\subfigure[]{\includegraphics[height=2.1in,width=2.26in]{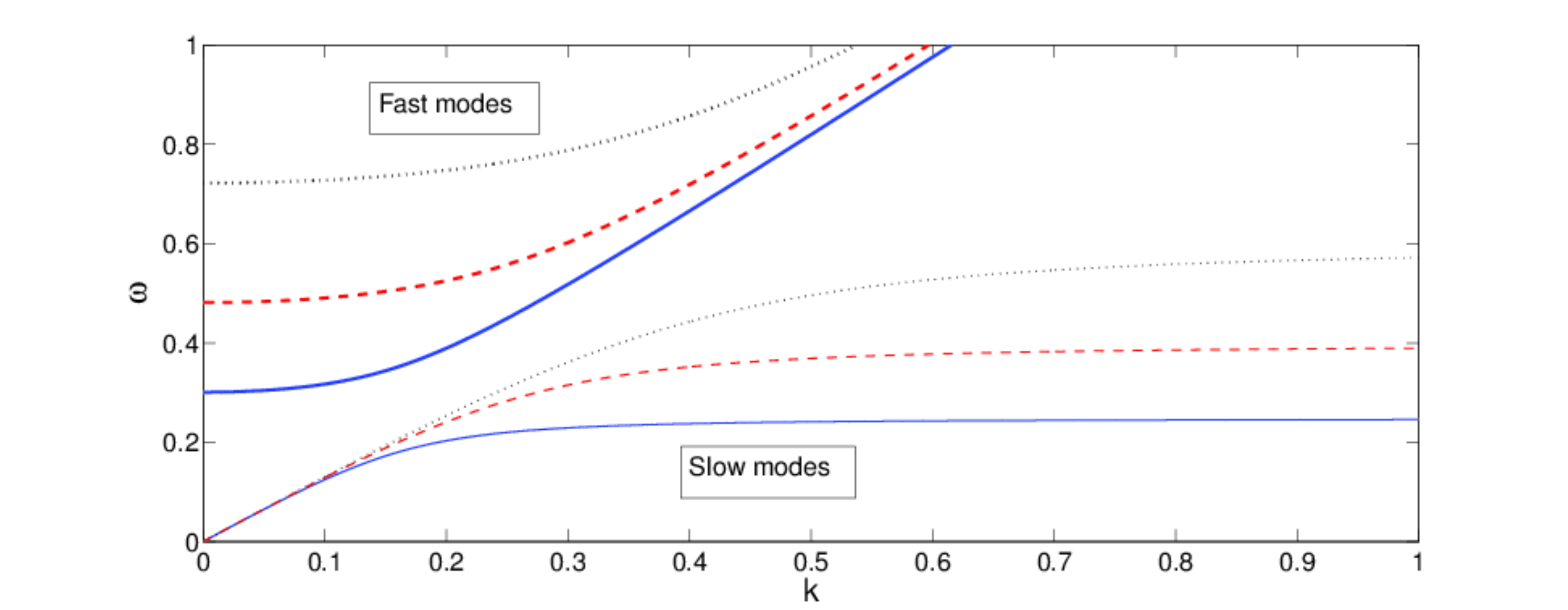}
\label{fig:subfigure1b}}
\subfigure[]{\includegraphics[height=2.1in,width=2.26in]{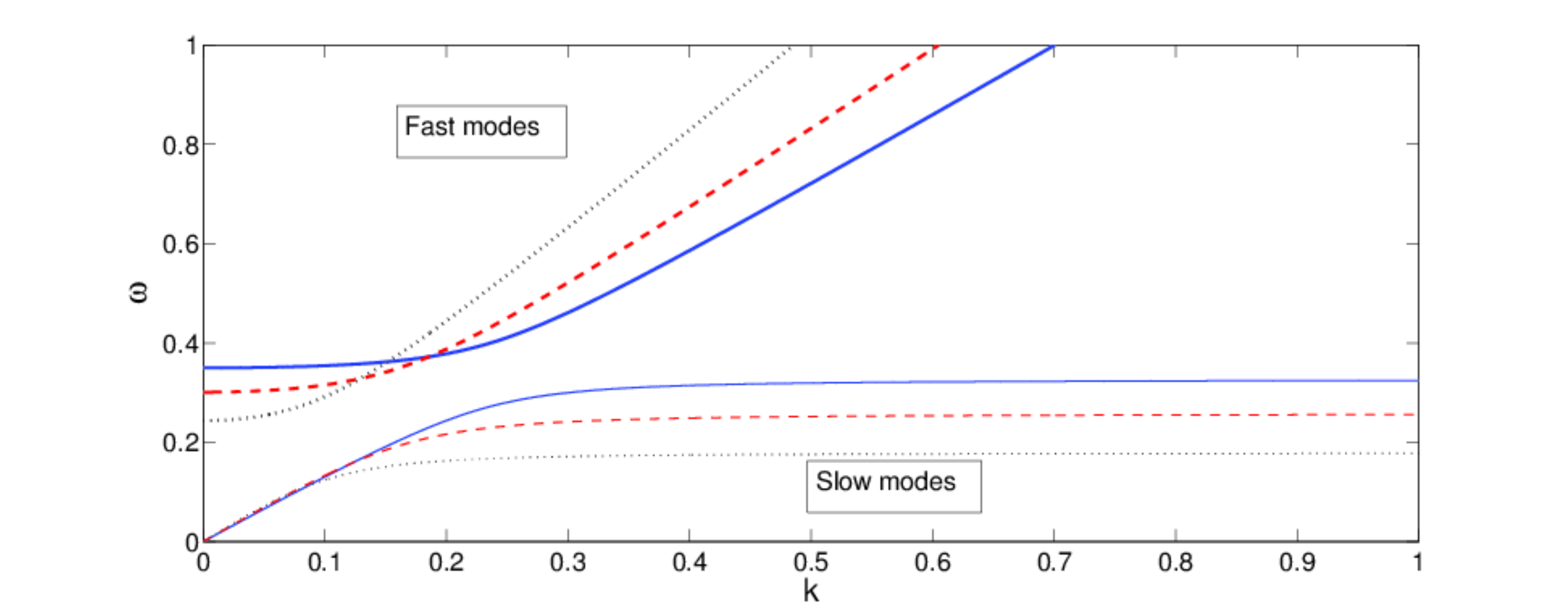}
\label{fig:subfigure1c}}
\subfigure[]{\includegraphics[height=2.1in,width=2.5in]{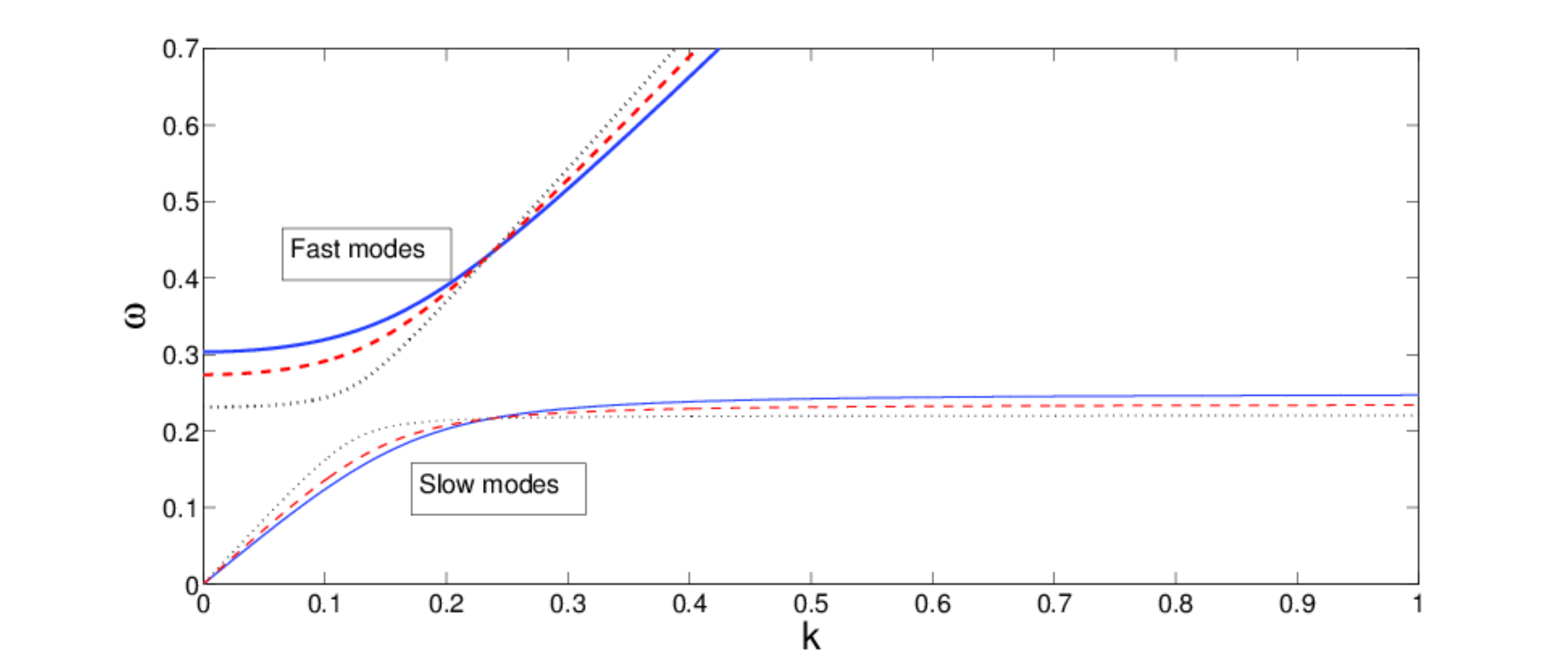}
\label{fig:subfigure1d}}
\subfigure[]{\includegraphics[height=2.1in,width=2.5in]{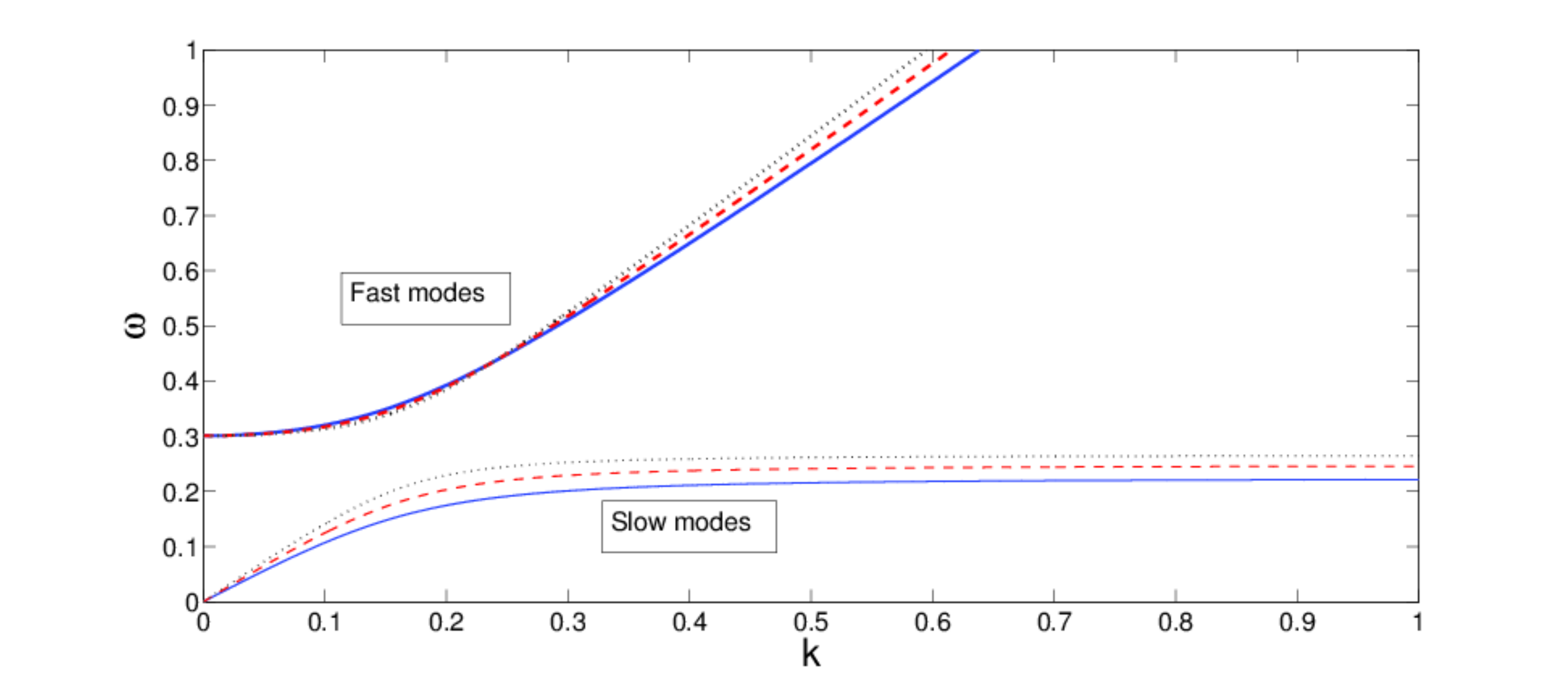}
\label{fig:subfigure1e}}
\caption{Effects of  (a) the obliqueness of propagation $(l_z)$, (b) the external magnetic field $(\omega_{cp})$, (c) the ratio of   negative and positive ion masses $(m)$  (d) positively charged dusts $(\delta)$ and  (e) the negative to positive ion temperature ratio $(T)$   on the low-frequency modes  given by Eq. \eqref{low-freq-mode} are shown.  In each subfigure, the upper  (three thick lines) and the lower (three thin lines) curves, respectively, correspond to the fast (DIC-like) and slow (DIA) modes. Subfigure (a): Different values of  the propagation angle $\theta$ are $\theta=\pi/6$ (solid line), $\pi/4$ (dashed line), and $\pi/3$ (dotted line); the   fixed parameter values are $T=0.7,~\delta=0.1,~m=3$, and $\omega_{cp}=0.5$.   Subfigure (b): Different values of $\omega_{cp}$ are  $\omega_{cp}=0.5$ (solid line), $\omega_{cp}=0.8$ (dashed line), and $\omega_{cp}=1.2$ (dotted line); the fixed parameter values are $T=0.5,~m=2$, $\theta=\pi/6$ and $\delta=0.1$. Subfigure (c): Different values of $m$ are  $m=1.5$ (solid line), $m=2.0$ (dashed line), and $m=3$ (dotted line); the fixed parameter values are $T=0.7$,  $\delta=0.1$, $\theta=\pi/6$ and $\omega_{cp}=0.5$.     Subfigure (d): Different values of $\delta$ are $\delta=0.05$ (solid line), $\delta=0.5$ (dashed line), and $\delta=0.9$ (dotted line); the fixed parameter values are $T=0.5,~m=2$, $\theta=\pi/6$  and $\omega_{cp}=0.5$. Subfigure (e): Different values of $T$ are  $T=0.1$ (solid line), $T=0.5$ (dashed line), and $T=0.9$ (dotted line); the fixed parameter values are $m=2$,  $\delta=0.1$, $\theta=\pi/6$ and $\omega_{cp}=0.5$. }
\label{fig:figure1}
\end{figure*}

  Figure \ref{fig:figure1} shows the characteristics of the low-frequency fast and slow modes [Eq. \eqref{low-freq-mode}] by the effects of (i) the obliqueness of wave propagation $(\theta)$, (ii) the static magnetic field  $(\omega_{cp})$,  (iii) the  negative to positive-ion mass ratio $(m)$, (iv) the immobile positively charged dusts $(\delta)$ and (v)  the negative to positive ion temperature ratio   $(T)$.  We find that  the obliquely propagating slow  (lower branch) and fast modes (upper branch) correspond, respectively, to the DIA and  DIC-like waves. Below we discuss the  properties of these modes separately.\\\\ 
(i) \textit{ Effect of oblique wave propagation:}  The effect of the propagation angle $\theta$ on the linear properties of the low-frequency modes  are shown in Fig. \ref{fig:subfigure1a}. It is found that increasing the angle with respect to the magnetic field leads to a decrease in the frequency $\omega$ of both the fast (upper branch) and slow (lower branch) modes along with the modification of their phase speeds. Furthermore,   the effect of $\theta$ on the fast mode is significant for $k\rightarrow0$, while its effect on the slow mode is noticeable for $k\gtrsim0.03$. It is also seen that the frequency gap between the modes decreases with increasing the angle $\theta$. An opposite trend of increasing the frequency gap by the effect of $\theta$ was found for oblique electron-acoustic waves in a magnetized kappa-distributed electron-ion plasma \cite{ppcf-sultana}.    \\\\  
(ii) \textit{Magnetic field effect:} The influence of the magnetic field strength (characterized by $\omega_{cj},~j=p,n$) on the wave modes is shown in Fig. \ref{fig:subfigure1b}. Both the weaker $(\omega_{cj}<1)$ and the stronger $(\omega_{cj}>1)$ magnetic fields increase  the frequency gap between the modes. However, a significant increase in the frequency of the DIC waves (upper branch), and hence a significant increase of the frequency gap between the modes, is seen to occur for stronger magnetic fields. It is also found that the phase speed of the DIA  modes remain constant as $k\rightarrow0$. These interesting features of DIA and DIC-like modes have not been observed before in pair-ion plasmas \cite{pair-ion1,pair-ion2,pair-ion3,pair-ion4,pair-ion5}\\\\
(iii) \textit{Effect of negative to positive ion mass ratio:} The effects of the mass ratio $m$  on the fast or DIC (upper branch) and slow or DIA (lower branch) modes are shown in Fig. \ref{fig:subfigure1c}. Interestingly, as $m$ increases, i.e., as the mass difference between the ions increases, the wave frequency  of the fast mode  increases for a wave number exceeding a critical value (This value is different for different values of $m$), otherwise it decreases as $k\rightarrow0$. On the other hand, the mass ratio $m$ has almost the similar influence on the slow modes  as  in Fig. \ref{fig:subfigure1b} (lower curves).\\\\
(iv) \textit{Effect of immobile charged dusts:}  Figure \ref{fig:subfigure1d} shows the effect of the presence of positively charged dusts in the background plasma.  It is found that increasing the concentration of charged dusts $\delta$ (hence decreasing the positive- to negative-ion density ratio $\mu$ to maintain the charge neutrality) leads to an increase in the frequency $\omega$ of the upper mode (DIC), and a frequency decrease in the lower DIA mode. This occurs  when the  wave number  exceeds its critical value $k_c$ lying in $0.2<k<0.3$. However, for $k<k_c$, the opposite trend occurs for both the modes along with the modification of their phase speeds. As a result,  the frequency gap between the modes for $k>k_c$ $(k<k_c)$  is found to increase  (decrease). Thus, the presence of immobile charged dusts not only modifies the wave frequency and the phase speed, but also reduces the  frequency gap between the DIA and DIC-like modes, which can not be observed   in pure pair-ion plasmas with equal mass and temperature \cite{pair-ion-experiment,saleem-pair-ion,pair-ion2,pair-ion3}.   However,  some different features of increasing   the frequency gap (not the decreasing trend) between DIA and DIC modes  were   found, e.g., for oblique propagation of electron-acoustic waves in a magnetized plasma without charged dusts \cite{ppcf-sultana}.   \\\\
(v) \textit{Effect of negative to positive ion temperature ratio:} The effects of the thermal pressures of ions on both the fast (upper branch) and slow (lower branch)  modes are exhibited in Fig. \ref{fig:subfigure1e}. It is seen that the temperature ratio $T$ of negative to positive ions has stronger influence with increasing the frequency of the DIA mode (lower branch) than the DIC (upper branch) mode. In the latter, the change of wave frequency with $T$ is noticeable for wave numbers satisfying  $k\gtrsim0.25$.  Furthermore,  as in Fig. \ref{fig:subfigure1d}, the increase (decrease) of the frequency gap between the modes   at short (long)-wavelengths is seen to occur. \\\\     

From the results (iii) and (v), one may thus conclude that the properties of the DIA and DIC modes in pair-ion plasmas with different mass and temperatures of ions are quite distinctive to those found in pair-ion plasmas with equal mass and temperature of ions \cite{pair-ion-experiment,saleem-pair-ion,pair-ion2,pair-ion3}. 
 
 \textit{Secondly}, for wave propagation along the magnetic filed we have $k_x$, $k_y\longrightarrow0$; $k_z=k_{\parallel}\neq0$. In this case, the transverse velocity components of the ion fluids   vanish [See Eqs. \eqref{velocity-x-p}, \eqref{velocity-y-p} and similar equations for negative ions, not shown], and particles will have velocities only along the magnetic field. Thus, we have   electrostatic DIA wave modes, the  dispersion relation of which is obtained from Eq. \eqref{disp-relation} as 
\begin{widetext}
\begin{equation}
\omega^2=\frac{1}{2}\left[1+m\mu+\frac{5}{3}(m+T)k^2_{\parallel}\pm\sqrt{\left(m\mu-1+\frac{5}{3}(m-T)k^2_{\parallel}\right)^2+4m\mu}\right]. \label{disp-parallel-1}
\end{equation}
\end{widetext}
The upper and lower signs in Eq. \eqref{disp-parallel-1}, respectively, correspond to the fast and slow DIA modes that are modified by the different mass and temperatures of ions as well as the presence of positively charged dusts. In particular, for plasmas with equal mass and temperature of ions, i.e., $m=T=1$, and with no dust, i.e., $\mu=1$, the slow wave becomes dispersionless with a phase speed independent of the wave number $k$, while the fast mode propagates similar to the high-frequency (in comparison with the plasma oscillation frequency) electron-acoustic  waves in an unmagnetized electron-ion plasma. 
In the quasineutrality limit (valid for long-wavelength modes), Eq. \eqref{disp-relation} reduces to
\begin{equation}
\omega=k_{\parallel}\sqrt{\frac53m\left(\frac{1+T\mu}{1+m\mu}\right)}. \label{disp-parallel-2}
\end{equation}

\begin{figure*}[ht]
\centering
\subfigure[]{\includegraphics[height=2.5in,width=3.3in]{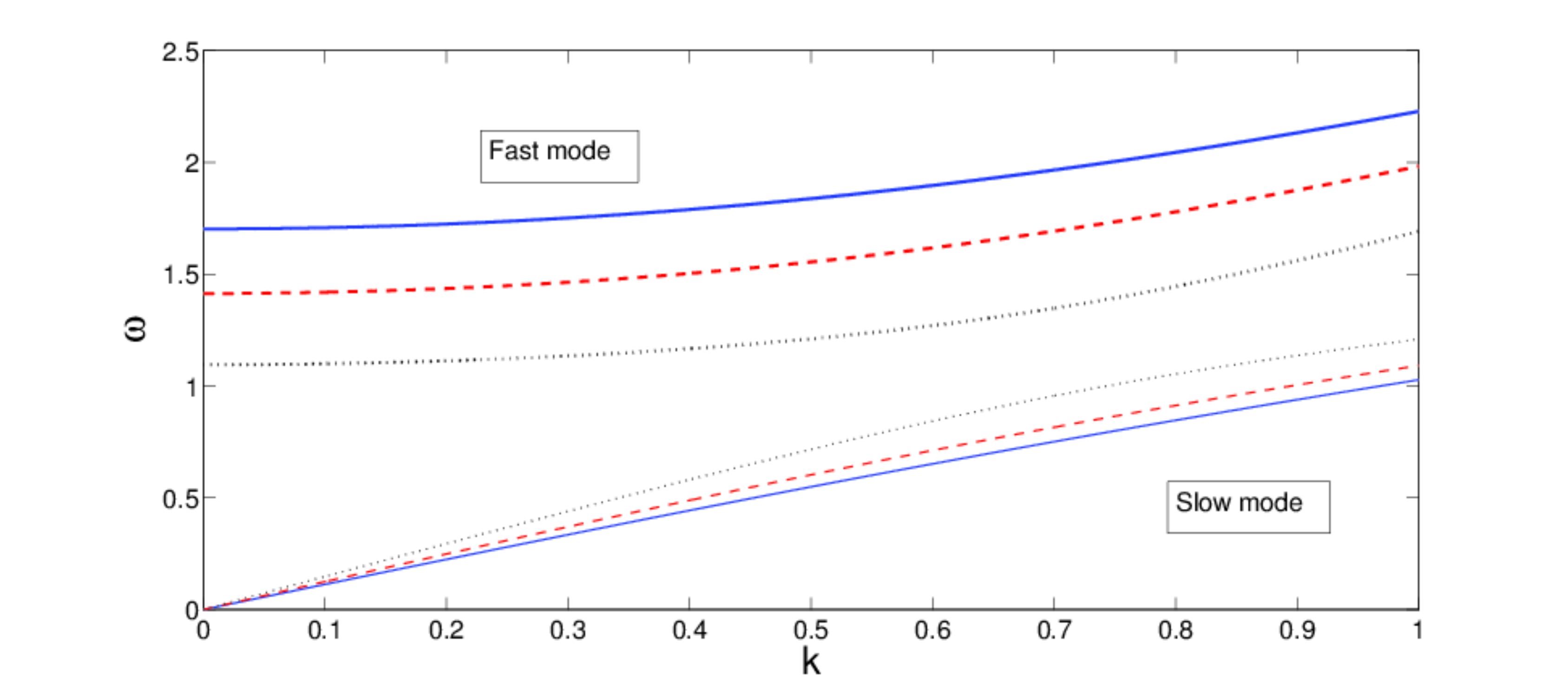}
\label{fig:subfigure2a}}
\quad
\subfigure[]{\includegraphics[height=2.5in,width=3.3in]{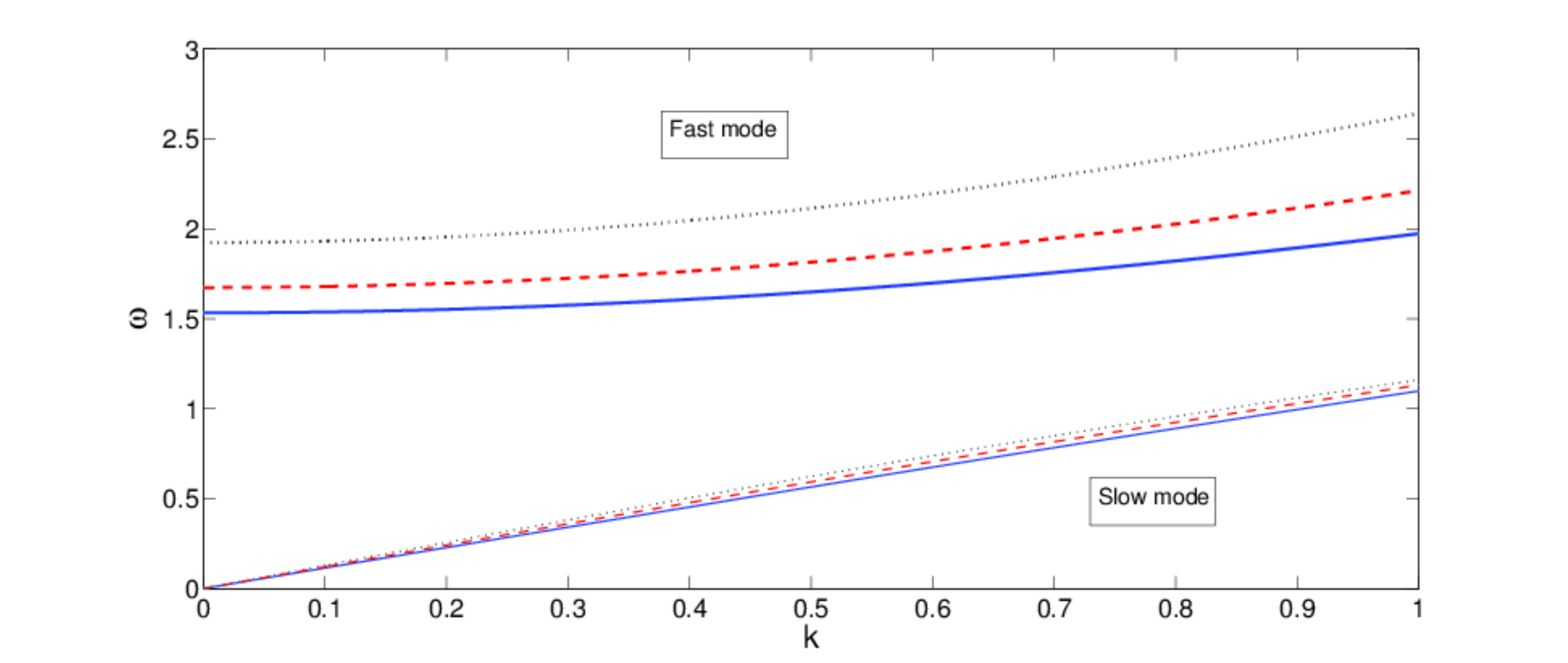}
\label{fig:subfigure2b}}
%
\caption{Effects of (a) positively charged dusts $(\delta)$     and (b) the ratio of   negative and positive ion masses $(m)$  on the DIA modes  given by Eq. \eqref{disp-parallel-1} are shown.  In each subfigure, the upper  (three thick lines) and the lower (three thin lines) curves, respectively, correspond to the fast   and slow   DIA modes. For the subfigures  (a) and (b), different values of the parameters $\delta$  and $m$,  as well as the corresponding fixed parameter values are considered as the same as in Figs. \ref{fig:subfigure1d} and \ref{fig:subfigure1c}  respectively.   }
\label{fig:figure2}
\end{figure*}
From Eqs. \eqref{disp-parallel-1} and \eqref{disp-parallel-2}, we find that the electrostatic waves, while propagating parallel to the static magnetic field,  are purely longitudinal acoustic-like waves (similar to the case of no magnetic field) with its frequency being independent of the magnetic field, and they are  dispersive due to the effects of charge separation (deviation from quasineutrality) of the ion fluids.   The phase velocities of these waves increase  with the wave number and are greater than the acoustic speed for the ion fluids. Furthermore,  these waves become dispersionless in the long-wavelength limits, and propagate with a constant phase speed.

 It may be instructive to analyze the properties of the DIA modes [Eq. \eqref{disp-parallel-1}], which typically depend  on the plasma parameters, namely $\delta,~T$ and $m$.  The features are shown in Fig. \ref{fig:figure2}.   Figure \ref{fig:subfigure2a} shows that  as the charged dust concentration increases (i.e., the positive to negative ion density ratio decreases), the frequency of the slow mode (lower branch) increases, while  for the fast modes it decreases significantly.  Furthermore, increasing the dust density leads to reducing the frequency gap between the modes. It is also seen that the effect  of $\delta$ is significant for both  the short and long-wavelength  fast modes, while its effect on the long-wavelength slow modes is comparatively weak.
  In contrast to the properties of the fast modes in  Fig. \ref{fig:subfigure2a} with the effect of $\delta$, Fig. \ref{fig:subfigure2b} shows that as the mass ratio $m$ increases, the frequencies of both the  fast and slow modes increase along with the increase in the frequency gap between the modes. It is evident from Eq. \eqref{disp-parallel-2} and from the lower branches of Fig. \ref{fig:figure2} that the phase velocity of the DIA mode remains constant as $k\rightarrow0$, i.e., in the long-wavelength limit. Furthermore, 
 it is found that (not shown in the figure) the effect  of the temperature ratio $T$  on the fast DIA modes is almost negligible, while it increases the frequency of the slow modes   similar to that observed in Fig. \ref{fig:subfigure2a}.

\textit{Lastly}, we consider the  wave propagation   perpendicular to the magnetic filed, i.e.,  propagation for $k_z\longrightarrow0$, $k^2_x+k^2_y=k^2_{\perp}\neq0$. In this case, ions  have  velocity components only transverse to the magnetic field.  The dispersion   Eq. \eqref{disp-relation} then reduces to
\begin{widetext}
\begin{equation}
\omega^2=\frac{1}{2}\left[1+m\mu+\omega^2_{cp}+\omega^2_{cn}+\frac{5}{3}(m+T)k^2_{\perp}\pm\sqrt{\left(m\mu-1+\omega^2_{cp}-\omega^2_{cn}+\frac{5}{3}(m-T)k^2_{\perp}\right)^2+4m\mu}\right]. \label{disp-perp-1}
\end{equation}
\end{widetext}
This represents a DIC  wave modified by the  effects of the external magnetic field,  different mass and temperature of the ion fluids as well   the presence of static charged dusts. Note that in the limit of vanishing-magnetic field, Eq. \eqref{disp-perp-1} trivially recovers the same form as Eq. \eqref{disp-parallel-1} for modified DIA waves.  In the quasineutrality limit, Eq. \eqref{disp-perp-1} reduces to the following  DIC mode  
\begin{equation}
\omega^2=\frac{\omega^2_{cp}+m\mu\omega^2_{cn}}{1+m\mu}+{\frac53m\left(\frac{1+T\mu}{1+m\mu}\right)k^2_{\perp}}. \label{disp-perp-2}
\end{equation}
In particular, for $m=T=1$  and $\omega_{cp}=\omega_{cn}=\omega_c$, we recover from Eq. \eqref{disp-perp-2}  the following ion-cyclotron mode similar to that appears in magnetized electron-ion plasmas with  Boltzmann distributed electrons:
\begin{equation}
\omega^2=\omega_c^2+\frac53k^2_{\perp}. \label{disp-perp-3}
\end{equation}
 The factor $5/3$ appears due to the adiabatic pressure of ion fluids in three-dimensional configuration.
 
\begin{figure*}[ht]
\centering
\subfigure[]{\includegraphics[height=2.1in,width=2.2in]{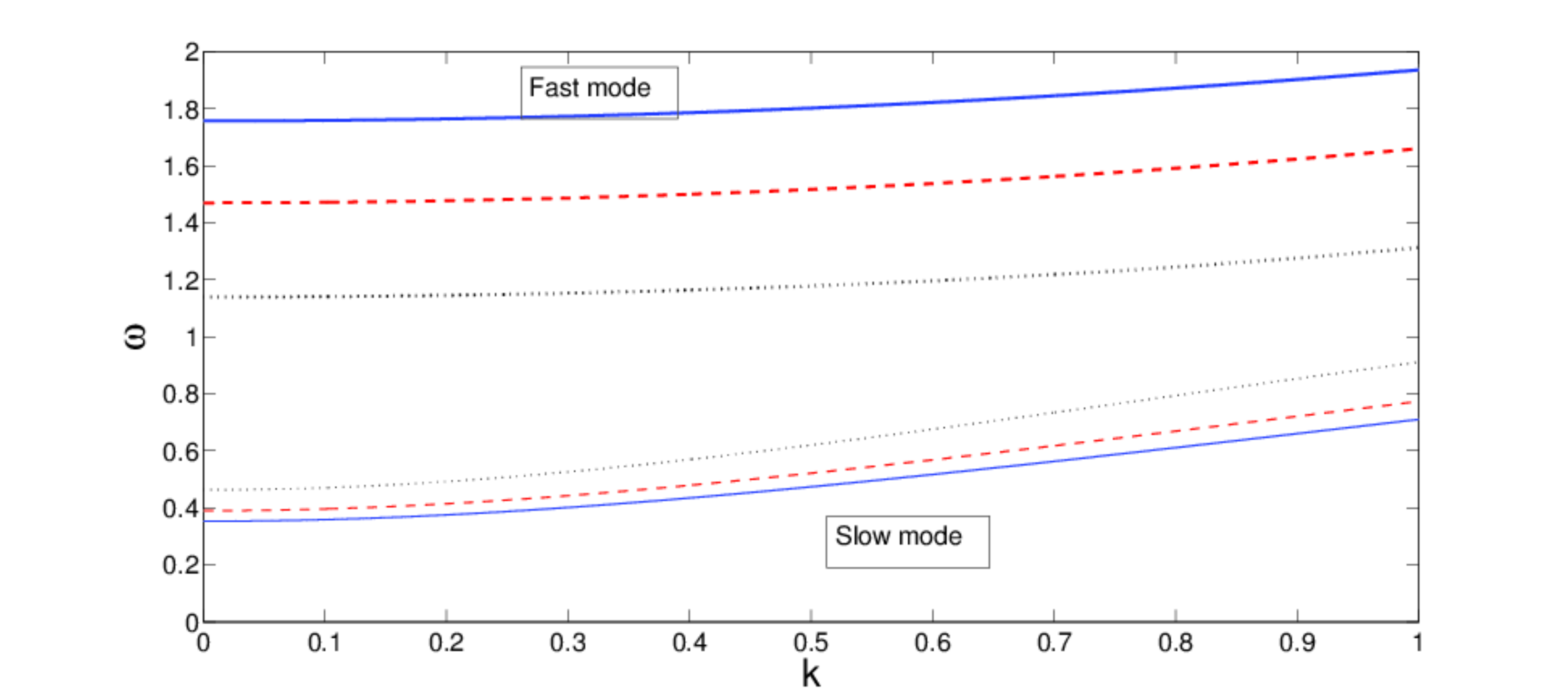}
\label{fig:subfigure3a}}
\subfigure[]{\includegraphics[height=2.1in,width=2.2in]{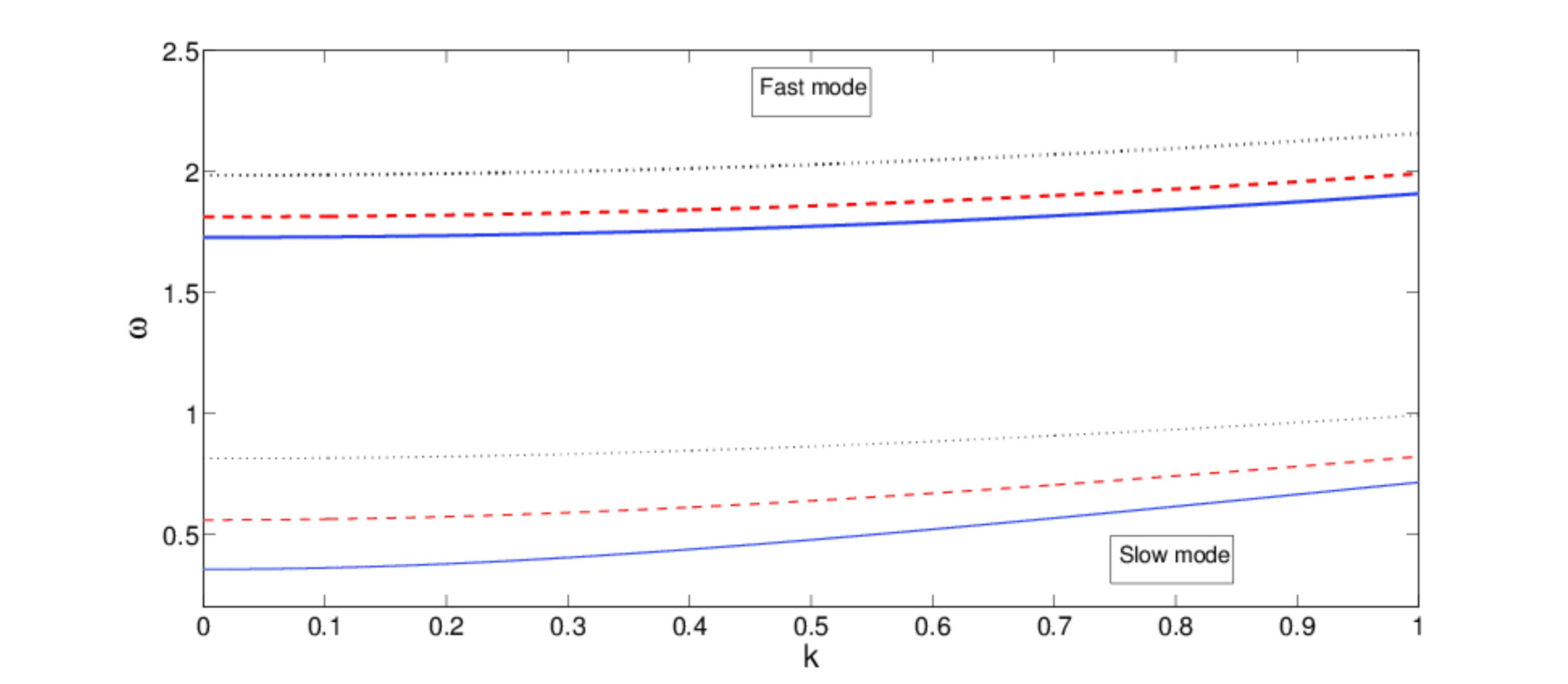}
\label{fig:subfigure3b}}
\subfigure[]{\includegraphics[height=2.1in,width=2.2in]{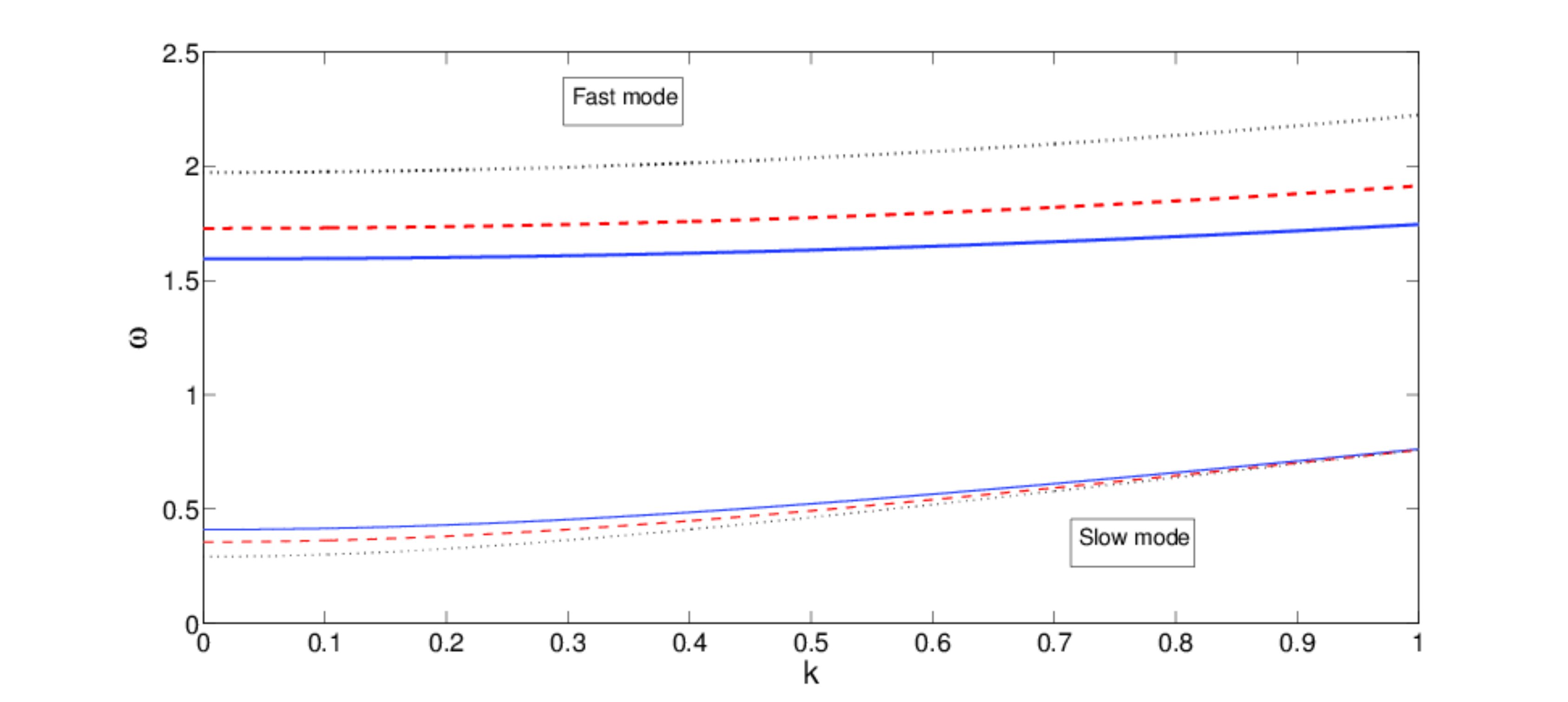}
\label{fig:subfigure3c}}
\caption{Effects of (a) positively charged dusts $(\delta)$,  (b) the external magnetic field $(\omega_{cp})$,   and (c) the ratio of   negative and positive ion masses $(m)$  on the DIC modes  given by Eq. \eqref{disp-perp-1} are shown.  In each subfigure, the upper  (three thick lines) and the lower (three thin lines) curves, respectively, correspond to the fast   and slow DIC   modes. For the subfigures  (a) to (c), different values of the parameters $\delta,~\omega_{cp}$, and $m$,  as well as the corresponding fixed parameter values are considered as the same as in Figs. \ref{fig:subfigure1d}, \ref{fig:subfigure1b} \ref{fig:subfigure1c} respectively.  }
\label{fig:figure3}
\end{figure*}

The properties  of the DIC modes given by Eq. \eqref{disp-perp-1} are shown in Fig. \ref{fig:figure3} for a set of parameters as in Fig. \ref{fig:figure1}. It is found that in contrast to the features of the slow mode (lower branch), as $\delta$ increases or  the density ratio $\mu$ decreases to maintain the quasineutrality, the frequency of the DIC mode decreases significantly [See Fig. \ref{fig:subfigure3a}]. The influence of the magnetic field (represented by $\omega_{cj}$) is depicted in Fig. \ref{fig:subfigure3b}. A stronger magnetic field increases the frequency of both the fast and slow modes.  In contrast to Fig. \ref{fig:subfigure3a},  Fig. \ref{fig:subfigure3c} shows that the effect of increasing the mass ratio $m$ is to increase (decrease)  the frequency of the fast (slow) modes. This change of frequency (increase) with different values of $m$ is significant for the fast modes, however, the slow wave frequency tends to approach  a constant value  for $k\gtrsim1$. In each of   the  subfigures, a significant modification in the wave phase speed  is seen to occur. Also,  as $k\rightarrow0$, the DIC frequency is found to remain almost constant with   $k$, while  the same increases   as $k\rightarrow1$.  Furthermore, it is found that (not shown in the figure) the ion temperatures do not have any significant effect on the fast mode as well as the long-wavelength slow modes. However, a significant   increase in the wave frequency of the slow modes for $k\gtrsim0.3$  is seen to occur for comparatively higher values of $T$. 
\section{Derivation of KdV equation}
In the linear analysis, we have neglected some interesting physics contained in the second-order perturbation terms like $n_{j1}v_{j1}$ etc. or   some higher-order terms. These are, indeed,   important when the wave grows in amplitude. In this section, we  are interested to consider  the nonlinear propagation (oblique to the magnetic field) of   small but finite amplitude  electrostatic DIA  waves   in a magnetized dusty pair-ion plasma. We follow the    standard reductive perturbation technique in which the stretched coordinates are defined as \cite{misra-samanta-jpp}
\begin{equation}
\begin{split}
\xi&=\epsilon^{1/2}({\hat{\kappa}}\cdot\mathbf{r}-Mt)\\
&=\epsilon^{1/2}(l_xx+l_yy+l_zz-Mt), \\
 &\tau=\epsilon^{3/2}t, \label{stretch-coord}
 \end{split}
\end{equation} 
 where $\epsilon$ is a small scaling parameter $(\lesssim1)$ measuring the weakness of perturbations and $M$ is the nonlinear wave speed  (relative to  the rest frame), normalized by $c_s$, to be determined later. Also,  $\hat{\kappa}$ is the unit vector along the direction of the wave propagation with $l_x$, $l_y$, $l_z$ denoting its direction cosines  along $x$, $y$ and $z$ axes respectively.   The dynamical variables are expanded as  \cite{misra-samanta-jpp}
\begin{equation}
\begin{split}
n_j&=1+\epsilon n_j^{(1)}+\epsilon^2 n_j^{(2)}+\cdots,\\
v_{jx,y}&=\epsilon^{3/2}v_{jx,y}^{(1)}+\epsilon^2v_{jx,y}^{(2)}+\cdots, \\
v_{jz}&=\epsilon v_{jz}^{(1)}+\epsilon^2v_{jz}^{(2)}+\cdots,  \\
\phi&=\epsilon\phi^{(1)}+\epsilon^2\phi^{(2)}+\cdots.\label{expansion-variables}
\end{split}
\end{equation}
Note that in the expansions \eqref{expansion-variables},   the first-order perturbations for the transverse velocity components of the ion fluids appear in  higher-orders of $\epsilon$ than that for the parallel components. For the nonlinear DIA waves, this anisotropy is introduced due to the fact that the ion gyro-motion (perpendicular to the magnetic field) is treated as a higher-order  effect than the motion parallel to the magnetic field \cite{mi-bains-pop,zk-misra-pop}.   

Next, we substitute  the expressions from Eqs. \eqref{stretch-coord} and \eqref{expansion-variables}  into   Eqs. \eqref{cont-eqn-nond}-\eqref{poisson-eqn-nond}, and equate different powers of $\epsilon$. 
 Thus, from Eq. \eqref{cont-eqn-nond}, equating successively the coefficients of  $\epsilon^{3/2}$, $\epsilon^2$ and  $\epsilon^{5/2}$, we obtain
 \begin{equation}
n^{(1)}_j=\frac{l_z}{M}v^{(1)}_{jz},\label{cont-perturb-n}
\end{equation}
\begin{equation}
v^{(1)}_{jx}=-\frac{l_y}{l_x}v^{(1)}_{jy},\label{cont-perturb-v}
\end{equation}
\begin{eqnarray}
&&-M\frac{\partial n^{(2)}_j}{\partial\xi}+\frac{\partial n^{(1)}_j}{\partial\tau}+l_x\frac{\partial v^{(2)}_{jx}}{\partial\xi}+l_y\frac{\partial v^{(2)}_{jy}}{\partial\xi}+l_z\frac{\partial v^{(2)}_{jz}}{\partial\xi}\notag \\
&&+l_z\frac{\partial\left({n^{(1)}_j} {v^{(1)}_{jz}}\right)}{\partial\xi}=0.\label{cont-perturb-n-v}
\end{eqnarray}
From the $x$ and $y$-components of   Eq. \eqref{moment-eqn-nond} for positive ions, 
 and equating the coefficients of $\epsilon^{3/2}$ and $\epsilon^2$, we successively obtain
\begin{equation}
0=-ml_{x,y}\frac{\partial\phi^{(1)}}{\partial\xi}\pm\omega_{cp} v^{(1)}_{p(y,x)}-\frac{5}{3}ml_{x,y}\frac{\partial n^{(1)}_p}{\partial\xi},\label{moment-xy-comp-p1}
\end{equation}
\begin{equation}
M\frac{\partial v^{(1)}_{p(x,y)}}{\partial\xi}=\mp\omega_{cp} v^{(2)}_{p(y,x)}.\label{moment-xy-comp-p2}
\end{equation}
Next, from the $z$-component of  Eq. \eqref{moment-eqn-nond} for positive ions, and
equating the coefficients of $\epsilon^{3/2}$ and $\epsilon^{5/2}$, we, respectively, obtain
\begin{equation}
v^{(1)}_{pz}=\frac{ml_z}{M}\left(\phi^{(1)}+\frac{5}{3} n^{(1)}_p\right),\label{moment-z-comp-p1}
\end{equation}
\begin{eqnarray}
&&-M\frac{\partial v^{(2)}_{pz}}{\partial\xi}+\frac{\partial v^{(1)}_{pz}}{\partial\tau}+l_z v^{(1)}_{pz}\frac{\partial v^{(1)}_{pz}}{\partial\xi}=\notag \\
&&ml_z\left[-\frac{\partial\phi^{(2)}}{\partial\xi}+\frac{5}{9}\left( n^{(1)}_p\frac{\partial n^{(1)}_p}{\partial\xi}-3\frac{\partial n^{(2)}_p}{\partial\xi}\right)\right].\label{moment-z-comp-p2}
\end{eqnarray}
Similarly, from the $x$ and $y$-components of  Eq. \eqref{moment-eqn-nond} for negative ions, and
equating the coefficients of $\epsilon^{3/2}$ and $\epsilon^2$, we successively obtain
\begin{equation}
0=l_{x,y}\frac{\partial\phi^{(1)}}{\partial\xi}\mp\omega_{cn} v^{(1)}_{n(y,x)}-\frac{5}{3}Tl_{x,y}\frac{\partial n^{(1)}_n}{\partial\xi},\label{moment-xy-comp-n1}
\end{equation}
\begin{equation}
M\frac{\partial v^{(1)}_{n(x,y)}}{\partial\xi}=\pm\omega_{cn} v^{(2)}_{n(y,x)}.\label{moment-xy-comp-n2}
\end{equation}
Also, from the $z$-component of  Eq. \eqref{moment-eqn-nond} for negative ions, and
equating the coefficients of $\epsilon^{3/2}$ and $\epsilon^{5/2}$, we obtain
\begin{equation}
 v^{(1)}_{nz}=-\frac{l_z}{M}\left(\phi^{(1)}-\frac{5}{3}T n^{(1)}_n\right),\label{moment-z-comp-n1}
\end{equation}
\begin{eqnarray}
&&-M\frac{\partial v^{(2)}_{nz}}{\partial\xi}+\frac{\partial v^{(1)}_{nz}}{\partial\tau}+l_z\left(v^{(1)}_{nz}\frac{\partial v^{(1)}_{nz}}{\partial\xi}-\frac{\partial\phi^{(2)}}{\partial\xi}\right)\notag \\
&&=\frac{5}{9}Tl_z\left( n^{(1)}_n\frac{\partial n^{(1)}_n}{\partial\xi}-3\frac{\partial n^{(2)}_n}{\partial\xi}\right).\label{moment-z-comp-n2}
\end{eqnarray}
Furthermore, from the  Poisson's equation, i.e., Eq. \eqref{poisson-eqn-nond}, and 
equating the coefficients of $\epsilon$ and $\epsilon^2$, we  successively have
\begin{equation}
n^{(1)}_n=\mu n^{(1)}_p,\label{poisson-comp1}
\end{equation}
\begin{equation}
\frac{\partial^2\phi^{(1)}}{\partial\xi^2}= n^{(2)}_n-\mu n^{(2)}_p.\label{poisson-comp2}
\end{equation}
\subsection*{First-order perturbations and nonlinear wave speed}
From Eqs. \eqref{cont-perturb-n}, \eqref{moment-z-comp-p1} and \eqref{moment-z-comp-n1}, after eliminating $v^{(1)}_{jz}$ we have 
\begin{equation}
n^{(1)}_j= \pm l^2_z\lambda_j\phi^{(1)},\label{n1-phi1}
\end{equation}
where the $\pm$ stand for positive $(j=p)$ and negative ions $(j=n)$. Also,   $\lambda_p={3m}/\left({3M^2-5ml^2_z}\right)$ and $\lambda_n=3/\left({3M^2-5Tl^2_z}\right)$. For typical plasma parameters with $m>1$ and $T<1$, the coefficients $\lambda_p$ and $\lambda_n$ are both positive for $M^2>5ml^2_z/3$. The latter will be validated  from the expression of $M$, to be obtained shortly. This  implies that the first-order density perturbations corresponding to the positive and negative ion fluids are of opposite in sign. The same also applies to the   perturbations of the velocity components of the ion fluids [See the expressions \eqref{vpn-z1-phi1} below].  
From Eqs. \eqref{moment-xy-comp-p1}, \eqref{moment-xy-comp-n1} and \eqref{n1-phi1}, after eliminating $n_j^{(1)}$, we obtain $(j=p,n)$
\begin{equation}
v^{(1)}_{j(x,y)}=\mp l_{y,x}\frac{M^2\lambda_j}{\omega_{cj}}\frac{\partial\phi^{(1)}}{\partial\xi},\label{vpn-xy1-phi1}
\end{equation}
where the $\mp$ stand for $x$ and $y$-components.
Also, from Eqs. \eqref{cont-perturb-n} and \eqref{n1-phi1}, we have 
\begin{equation}
v^{(1)}_{jz}=\pm Ml_z\lambda_j\phi^{(1)}.\label{vpn-z1-phi1}
\end{equation}
We note that Eq. \eqref{vpn-xy1-phi1} satisfies Eq. \eqref{cont-perturb-v} as required. Then from Eqs. \eqref{poisson-comp1} and \eqref{n1-phi1}, we obtain the expression for the wave speed in the moving frame of reference as
\begin{equation}
M=l_z\sqrt{\frac{5}{3}m\left(\frac{1+\mu T}{1+\mu m}\right)}.\label{phase-velo}
\end{equation}
From the expression of $\lambda_p$, it is clear that  $M^2\neq{5}m l^2_z/3$, so $T\neq m$, i.e., the temperature ratio and the mass ratio can not assume the same value. This implies that the present nonlinear theory may not be valid for  plasmas with equal mass and temperature of different ion fluids (i.e., the case of $M=T=1$). However, one may consider for laboratory and space plasmas (e.g., as in  Ref. \cite{pair-ion5}) that $T<1$ and $m>1$ for which    $M^2>{5}m l^2_z/3$. The latter is satisfied for   $\lambda_j>0$ as mentioned before. Note that Eq. \eqref{phase-velo}   represents the phase speed of the  obliquely propagating DIA wave in the $\xi-\tau$ frame of reference, which may correspond to the low-frequency long-wavelength slow mode given by Eq. \eqref{low-freq-mode}.   Interestingly, for $l_z=1$, i.e., for  wave propagation parallel to the magnetic field,  Eq. \eqref{phase-velo} becomes exactly the same as  Eq. \eqref{disp-parallel-2}.  We also find that the value of $M$ can be either $>1$ or $<1$ depending on the plasma parameters we consider. For example, for laboratory plasmas as in Table \ref{table:laboratory-plasma-parameters}, i.e., for $m=3.74$, $T=0.125$, $\mu=0.65$, we can obtain $M=0.7$ or $M=1.12$ when $l_z=0.5$  or $l_z=0.8$.   On the other hand, for space plasma environments as in Table \ref{table:space-plasma-parameters}, i.e., for $m=10.7$, $T=1$, $\mu=0.5$, we have $M=0.41$ and $1.03$ for $l_z=0.2$ and $0.5$ respectively.  It is seen that increasing (decreasing) the values of the parameters leads to an increase (decrease) in the value of $M$.
\subsection*{Second-order perturbations}
From Eqs. \eqref{moment-xy-comp-p2} and  \eqref{moment-xy-comp-n2}   using Eq. \eqref{vpn-xy1-phi1}  we  obtain
\begin{equation}
v^{(2)}_{j(x,y)}=\pm l_{x,y}\frac{ M^3\lambda_j}{\omega^2_{cj}}\frac{\partial^2\phi^{(1)}}{\partial\xi^2},\label{v2pnxy}
\end{equation}
where the $\pm$ stand for positive $(j=p)$ and negative ions $(j=n)$.
Next, from Eq. \eqref{cont-perturb-n-v} we eliminate $v_j^{(2)}$ using Eq. \eqref{v2pnxy},   and use Eq. \eqref{cont-perturb-n} to substitute the expressions for $n_j^{(1)}$ in terms of $\phi_j^{(1)}$ in the resulting equation. Thus, we obtain for positive and negative ions as
\begin{eqnarray}
&&\frac{\partial n^{(2)}_j}{\partial\xi}\mp\lambda_jl^2_z\frac{\partial\phi^{(2)}}{\partial\xi}=\pm\lambda_j^2\frac{Mm_j}{m_n}\left[M^2\frac{\left(1-l^2_z\right)}{\omega^2_{cj}}\frac{\partial^3\phi^{(1)}}{\partial\xi^3}\right.\notag\\
&&\left.+2l_z^2\frac{\partial\phi^{(1)}}{\partial\tau}\pm\lambda_jMl^4_z\left(27-\frac{5m_nT_jl^2_z}{m_jT_pM^2}\right)\phi^{(1)}\frac{\partial\phi^{(1)}}{\partial\xi}\right], \label{npn2-phi}
\end{eqnarray}
where the upper (lower) sign stands for $j=p~(j=n)$. 
\subsection*{KdV equation}
Substituting the expressions for $\partial n_j^{(2)}/\partial\xi$, $j=p,n$, to be obtained from Eq. \eqref{npn2-phi}  into Eq. \eqref{poisson-comp2}, and noting that coefficient of $\phi^{(2)}$ vanishes by Eq. \eqref{phase-velo},  we obtain the following  KdV   equation
\begin{equation}
\frac{\partial\Phi}{\partial\tau}+A\Phi\frac{\partial\Phi}{\partial\xi}+B\frac{\partial^3\Phi}{\partial\xi^3}=0,\label{K-dV}
\end{equation}
where $\Phi\equiv\phi^{(1)}\lesssim1$, such that $\epsilon\phi^{(1)}<1$. The coefficients of nonlinearity and dispersion are,  respectively, given by
\begin{eqnarray}
A&&=\frac{1}{2\sqrt 15}\frac{l_z}{(m-T)}\sqrt{\frac{m}{(1+\mu m)(1+\mu T)}}\notag \\
&&\times\left[\left(1+\mu^2 T\right)\left(1+\mu m\right)-9\left(1+\mu^2 m\right)\left(1+\mu T\right)\right],\label{coeff-dispersion-A}\notag \\ 
\end{eqnarray}
\begin{eqnarray}
B&&=\frac{5\sqrt5}{6\sqrt3}l_z\sqrt{m}\frac{(1+\mu T)^{3/2}}{(1+\mu m)^{5/2}}\notag \\
&&\times\left[\frac{\mu(m-T)^2}{(1+\mu T)^2}+\frac{(1-l^2_z)(1+m^3\mu)}{\omega_{cp}\omega_{cn}}\right].\label{coeff-nonlinear-B}
\end{eqnarray}
\begin{figure}[ht]
\centering
\includegraphics[height=2.0in,width=3.5in]{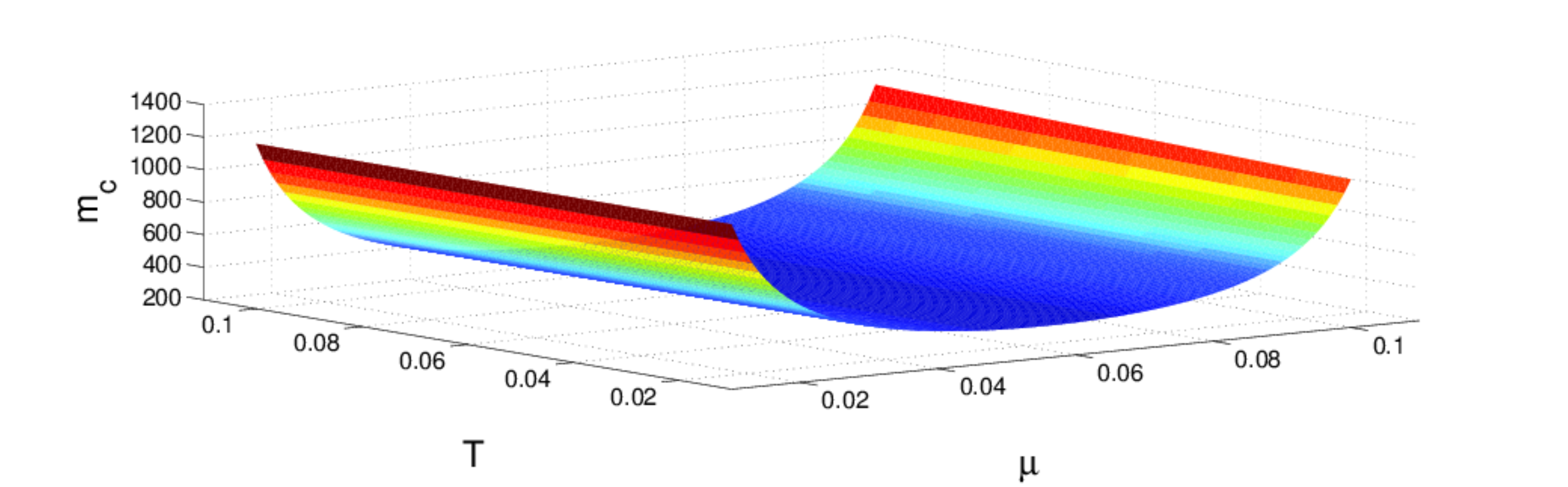}
\caption{Typical variations of the critical mass $m_c$ are shown with respect to the density and the temperatures ratios $\mu$ and $T$.  }
\label{fig:figure4}
\end{figure}

Inspecting on the expressions for $A$ and $B$ we find that $B$ is always positive.  However, $A\gtreqqless0$  according to when $m \gtreqqless m_c$, where $m_c$ is a critical value of $m$ given by
\begin{equation}
m_c=\frac{8+\mu T(9-\mu)}{\mu\left[1-\mu(9+8\mu T)\right]}. \label{m-critical}
\end{equation} 
In order that $m_c>0$, we must have $T<T_c\equiv(1-9\mu)/8\mu^2<1$ and $\mu<1/9$. For typical plasma parameters (See Tables \ref{table:laboratory-plasma-parameters} and \ref{table:space-plasma-parameters}) with $m>1$ and $T<T_c,\mu<1/9$, the values of $m_c$ become much larger than unity (e.g., for $\mu=T=0.1$, we have $m_c=879.2$), and so the dispersion coefficient $A$ is always negative for $m\ll m_c$. Thus, typical laboratory and space plasmas as in Tables \ref{table:laboratory-plasma-parameters} and \ref{table:space-plasma-parameters} with a pair of ions may support only rarefactive DIA solitons.  On the other hand, for $m\sim m_c$ the KdV equation \eqref{K-dV} fails to describe the evolution of DIA waves. In this case, one has go for higher-order corrections to derive a modified KdV equation which may admit a compressive DIA soliton solution in plasmas with $m\gg1$, i.e., with much heavier negative ions than positive ones \cite{rosenberg2009}.   Typical variations of $m_c$ with respect to $\mu$ and $T$ are shown in Fig. \ref{fig:figure4}. It is clear that  for a fixed $\mu$ as  $T$ increases, the value of $m_c$ increases. However, for a fixed $T$, $m_c$ decreases (increases) in the subinterval $0.01\lesssim\mu\lesssim0.07$ $(0.07<\mu\lesssim0.11)$.  Furthermore, the case of $m\gg m_c$ for which $A>0$ may not be relevant for the present study.   We also note that this critical value of $m$ may not appear in a magnetized pair-ion plasma  without stationary charged dusts, because in this case, the term in the square brackets in the expression of $A$ becomes $-8(1+m)(1+T)$, i.e.,   negative. Thus, in magnetized pure pair-ion plasmas, there may exist DIA solitary waves with only the negative potential.

A stationary solitary solution of Eq. \eqref{K-dV}   can be obtained by applying a transformation $\eta=\xi-U_0\tau$ to Eq. \eqref{K-dV}, where $U_0$ is the constant phase speed normalized by $c_s$, and imposing the boundary conditions for localized perturbations, namely, $\phi$, $d\phi/d\xi$, $d^2\phi/d\xi^2\rightarrow0$ as $\xi\rightarrow\pm\infty$, as  
\begin{equation}
\Phi=\Phi_m\text{sech}^2(\eta/w), \label{small-soliton}
\end{equation}
where $\Phi_m=3U_0/A$ is the finite amplitude and $w=\sqrt{4B/U_0}$ is the finite width of the soliton with $B>0$.    

\begin{figure*}[ht]
\centering
\includegraphics[height=4in,width=6in]{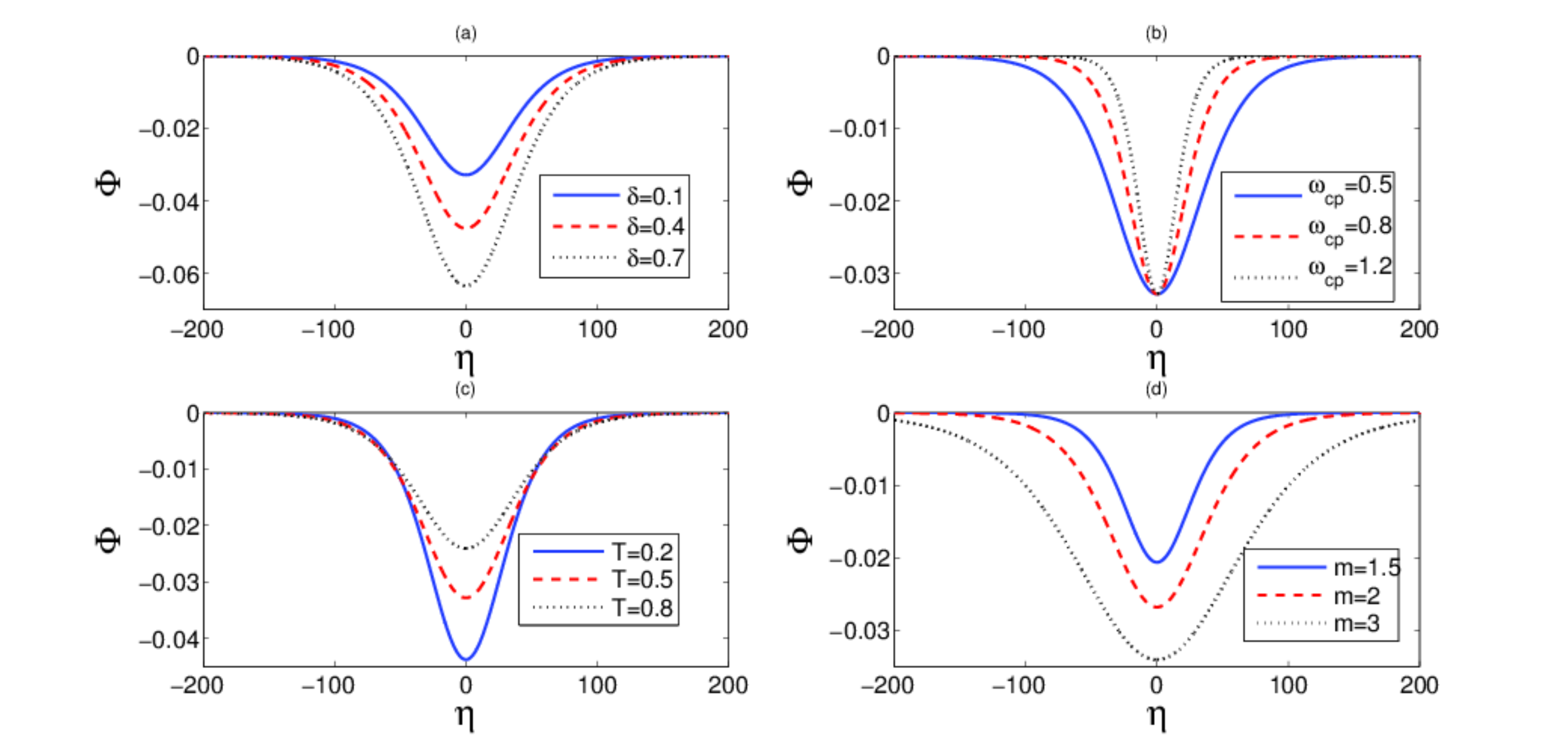}
\caption{Stationary soliton solution [Eq. \eqref{small-soliton}] of the KdV equation \eqref{K-dV} at $\tau=0$, and with $U_0=0.2$ is plotted against $\eta$. Subplots (a) to (d), respectively, show the effects of (a) positively charged dusts $(\delta)$,  (b) the external magnetic field $(\omega_{cp})$, (c) the   negative to positive ion temperature ratio $(T)$, and (d) the ratio of   negative and positive ion masses $(m)$.   The corresponding fixed parameter values are  as in subfigures. \ref{fig:subfigure1d}, \ref{fig:subfigure1b}, \ref{fig:subfigure1e} and \ref{fig:subfigure1c} respectively. }
\label{fig:figure5}
\end{figure*}

Relying on the   coefficients $A$ and $B$ of the KdV equation \eqref{K-dV},    we study the     properties of  the   rarefactive solitons   \eqref{small-soliton} (See Fig. \ref{fig:figure5}) for different values of the parameters as in Fig. \ref{fig:figure1}.   It is found that, for   fixed values of the mass ratio $m=2$, the temperature ratio $T=0.5$, and the magnetic field given by $\omega_{cp}=0.5$, as the charged dust concentration increases, both the amplitude and width of the soliton increases.    Figure \ref{fig:figure5}(b) shows that    the  magnetic field has influence only on the width of the soliton, since the effect of $B_0$ is only entered into the dispersion coefficient $B$. A stronger magnetic field  reduces the width, while the amplitude remains constant.   The influences of the temperature ratio $(T)$ on the amplitude and width of the soliton are shown in Fig. \ref{fig:figure5}(c). It is found that  the amplitude decreases significantly, while the width increases as the value of $T$ increases. A significant decrease in the amplitude as well as a significant increase in the width of the solitons are found with increasing values of the mass ratio  $m$ [Fig. \ref{fig:figure5}(d)]. Furthermore, since both the nonlinear and dispersion coefficients $A$ and $B$ are proportional to the obliqueness parameter $l_z$, its effect  is to decrease   (increase) the  amplitude (width) of the DIA soliton.
\section{Derivation of mKdV equation}
In the previous section we noticed that when $m=m_c$, the nonlinear coefficient of the  KdV equation \eqref{K-dV} vanishes, i.e., $A=0$. In this situation, Eq. \eqref{K-dV} fails to describe the evolution of small-amplitude DIA waves.  Thus, for values of $m\sim m_c$, we derive a modified KdV equation using the same reductive perturbation technique as in the previous section.  We  also take the same stretched coordinates for space   and time. However, the dependent variables are expanded in a different manner as  
\begin{equation}
\begin{split}
n_j&=1+\epsilon^{1/2} n_j^{(1)}+\epsilon n_j^{(2)}+\epsilon^{3/2} n_j^{(3)}+\cdots,  \\
v_{jx,y}&=\epsilon v_{jx,y}^{(1)}+\epsilon^{3/2} v_{jx,y}^{(2)}+\epsilon^2 v_{jx,y}^{(3)}+\cdots, \\
v_{jz}&=\epsilon^{1/2} v_{jz}^{(1)}+\epsilon v_{jz}^{(2)}+\epsilon^{3/2} v_{jz}^{(3)}+\cdots,  \\
\phi &=\epsilon^{1/2} \phi^{(1)}+\epsilon \phi^{(2)}+\epsilon^{3/2} \phi^{(3)}+\cdots.\label{expansion-mkdv}  
\end{split}
\end{equation}

We substitute the  expansions \eqref{expansion-mkdv} and the stretched coordinates \eqref{stretch-coord} into   Eqs. \eqref{cont-eqn-nond}-\eqref{poisson-eqn-nond}, and equate different powers of $\epsilon$ successively. 
In the lowest orders of $\epsilon$, i.e., $\epsilon$ and $\epsilon^{1/2}$ we obtain the same expressions for the first-order perturbations and the nonlinear wave speed $M$ as Eqs. \eqref{n1-phi1}-\eqref{phase-velo}.  However, from Eq. \eqref{cont-eqn-nond} equating the coefficients    of $\epsilon^{3/2}$ and $\epsilon^2$, we successively  obtain   
\begin{equation}
M n^{(2)}_{j}-l_zv^{(2)}_{jz}=l_xv^{(1)}_{jx}+l_yv^{(1)}_{jy}+l_zn^{(1)}_{j}v^{(1)}_{jz},\label{cont-2-mkdv}
\end{equation}
\begin{eqnarray}
&&-M\frac{\partial n^{(3)}_j}{\partial\xi}+\frac{\partial n^{(1)}_j}{\partial\tau}+l_x\frac{\partial} {\partial\xi}\left(n^{(1)}_{j}v^{(1)}_{jx}+v^{(2)}_{jx}\right)\notag \\
&&+l_y\frac{\partial} {\partial\xi}\left(n^{(1)}_{j}v^{(1)}_{jy}+v^{(2)}_{jy}\right)\notag \\
&&+l_z\frac{\partial} {\partial\xi}\left(n^{(1)}_{j}v^{(2)}_{jz}+n^{(2)}_{j}v^{(1)}_{jz}+v^{(3)}_{jz}\right)=0.\label{cont-3mkdv}
\end{eqnarray}
Next, from the $x$ and $y$-components of Eq. \eqref{moment-eqn-nond} for positive ions, and equating the coefficients of   $\epsilon^{3/2}$, we have
\begin{eqnarray}
M\frac{\partial v^{(1)}_{p(x,y)}}{\partial\xi}&&=ml_{x,y}\frac{\partial\phi^{(2)}}{\partial\xi}\mp\omega_{cp} v^{(2)}_{p(y,x)}+\frac53 ml_{x,y}\frac{\partial n^{(2)}_p}{\partial\xi}\notag \\
&&-\frac{5}{18}ml_{x,y}\frac{\partial (n^{(1)}_p)^2}{\partial\xi},\label{moment-2p-xy-mkdv}
\end{eqnarray}
where the $\mp$ stand for the  $x$ and $y$-components respectively.
Furthermore, from  the $z$-component of   Eq. \eqref{moment-eqn-nond} for positive ions, equating the coefficients of  $\epsilon^{3/2}$ and $\epsilon^2$  we have
\begin{eqnarray}
&&-Mv^{(2)}_{pz}+\frac{l_z}{2}(v^{(1)}_{pz})^2=-ml_z\phi^{(2)}\notag \\
&&+\frac{5}{3}ml_z\left[\frac16\left(n^{(1)}_p\right)^2-n^{(2)}_p\right],\label{moment-2p-z-mkdv}
\end{eqnarray}
 \begin{eqnarray}
&&-M\frac{\partial v^{(3)}_{pz}}{\partial\xi}+\frac{\partial v^{(1)}_{pz}}{\partial\tau}+\left(l_xv^{(1)}_{px}+l_yv^{(1)}_{py}\right)\frac{\partial v^{(1)}_{pz}}{\partial\xi}\notag \\
&&+l_x\frac{\partial(v^{(1)}_{pz}v^{(2)}_{pz})}{\partial\xi}=-ml_z\frac{\partial\phi^{(3)}}{\partial\xi}\notag \\
&&+\frac59 ml_z\left[\frac{\partial (n^{(1)}_pn^{(2)}_p)}{\partial\xi}-3\frac{\partial n^{(3)}_p}{\partial\xi}-\frac{2}{9}\frac{\partial(n^{(1)}_p)^3}{\partial\xi}\right].\label{moment-3p-z-mkdv}
\end{eqnarray}

Similar expressions for the negative ions can also be obtained.   Thus, from the $x$ and $y$-components of Eq. \eqref{moment-eqn-nond}, equating the  coefficients of $\epsilon^{3/2}$, we successively obtain
\begin{eqnarray}
&&-M\frac{\partial v^{(1)}_{n(x,y)}}{\partial\xi}=l_{x,y}\frac{\partial\phi^{(2)}}{\partial\xi}\mp\omega_{cn} v^{(2)}_{n(y,x)}-\frac53 Tl_{x,y}\frac{\partial n^{(2)}_n}{\partial\xi}\notag \\
&&+\frac{5}{18}Tl_{x,y}\frac{\partial (n^{(1)}_p)^2}{\partial\xi},\label{moment-2n-xy-mkdv}
\end{eqnarray}
where the $\mp$ stand for the  $x$ and $y$-components respectively.
From the $z$-component of   Eq. \eqref{moment-eqn-nond}   equating the coefficients of   
  $\epsilon^{3/2}$ and $\epsilon^2$  we have
\begin{equation}
-\frac{M}{l_z} v^{(2)}_{nz}+\frac{1}{2} \left(v^{(1)}_{nz}\right)^2=\phi^{(2)}+\frac{5}{3}T\left[\frac16\left(n^{(1)}_n\right)^2-n^{(2)}_n\right],\label{moment-2n-z-mkdv}
\end{equation}
 \begin{eqnarray}
&&-M\frac{\partial v^{(3)}_{nz}}{\partial\xi}+\frac{\partial v^{(1)}_{nz}}{\partial\tau}+\left(l_xv^{(1)}_{nx}+l_yv^{(1)}_{ny}\right)\frac{\partial v^{(1)}_{nz}}{\partial\xi}\notag \\
&&+l_x\frac{\partial(v^{(1)}_{nz}v^{(2)}_{nz})}{\partial\xi}=ml_z\frac{\partial\phi^{(3)}}{\partial\xi}\notag \\
&&+\frac59 Tl_z\left[\frac{\partial (n^{(1)}_nn^{(2)}_n)}{\partial\xi}-3\frac{\partial n^{(3)}_n}{\partial\xi}-\frac{2}{9}\frac{\partial(n^{(1)}_n)^3}{\partial\xi}\right].\label{moment-3n-z-mkdv}
\end{eqnarray}
Furthermore, from   Eq. \eqref{poisson-eqn-nond} equating the coefficients of $\epsilon$ and $\epsilon^{3/2}$ we obtain
\begin{equation}
n^{(2)}_n=\mu n^{(2)}_p,\label{poisson-2-mkdv}
\end{equation}
\begin{equation}
\frac{\partial^2\phi^{(1)}}{\partial\xi^2}=n^{(3)}_n-\mu n^{(3)}_p.\label{poisson-3-mkdv}
\end{equation}
\subsection*{Second-order perturbations}
From Eqs. \eqref{cont-2-mkdv},  \eqref{moment-2p-z-mkdv} and \eqref{moment-2n-z-mkdv}  solving for $n^{(2)}_j$ and   $v^{(2)}_{jz}$ $(j=p,n)$, and  substituting the corresponding   first-order quantities from Eqs. \eqref{n1-phi1} and \eqref{vpn-xy1-phi1} wherever necessary,  we obtain
\begin{equation}
n^{(2)}_j=\zeta_j\lambda_jl_z^2\phi^{(2)}+\frac{m_j}{2m_n}\lambda^3_jl_z^4\left(3M^2-\frac59 \frac{m_nT_j}{m_jT_p}l^2_z\right)(\phi^{(1)})^2, \label{npn2-mkdv}
\end{equation}
\begin{eqnarray}
v^{(2)}_{jz}&&=3\zeta_jM\lambda_jl_z\phi^{(2)}\notag\\
&&+\frac{M}{18}\frac{m_j}{m_n}\lambda^3_jl_z^3\left(9M^2+25 \frac{m_nT_j}{m_jT_p}l^2_z\right)(\phi^{(1)})^2, \label{vpnz2-mkdv}
\end{eqnarray}
where $\zeta_p=1$ and $\zeta_n=-1$.
Similarly, we solve for $v^{(2)}_{jx}$ and $v^{(2)}_{jy}$, $j=p,n$, from Eqs. \eqref{moment-2p-xy-mkdv} and \eqref{moment-2n-xy-mkdv} to obtain
\begin{eqnarray}
v^{(2)}_{jx}=&&-\frac{m_nl_y}{m_j\omega_{cj}}\left(1+\frac{5T_j}{3T_p}l_z^2\lambda_j\right)\frac{\partial\phi^{(2)}}{\partial\xi}+\zeta_jM^3l_x\frac{\lambda_j}{\omega^2_{cj}}\frac{\partial^2\phi^{(1)}}{\partial\xi^2}\notag \\
&&-\zeta_j\frac{5l_y}{54}\frac{T_j}{T_p}\frac{l_z^4\lambda^2_j}{\omega_{cj}}\left[\lambda_j\left(27M^2-5\frac{m_nT_j}{m_jT_p}l_z^2\right)-3\frac{m_n}{m_j}\right]\notag\\
&&\times\frac{\partial(\phi^{(1)})^2}{\partial\xi},\label{vpnx2-mkdv}
\end{eqnarray}
\begin{eqnarray}
v^{(2)}_{jy}=&&\frac{m_nl_x}{m_j\omega_{cj}}\left(1+\frac{5T_j}{3T_p}l_z^2\lambda_j\right)\frac{\partial\phi^{(2)}}{\partial\xi}+\zeta_jM^3l_y\frac{\lambda_j}{\omega^2_{cj}}\frac{\partial^2\phi^{(1)}}{\partial\xi^2}\notag \\
&&+\zeta_j\frac{5l_x}{54}\frac{T_j}{T_p}\frac{l_z^4\lambda^2_j}{\omega_{cj}}\left[\lambda_j\left(27M^2-5\frac{m_nT_j}{m_jT_p}l_z^2\right)-3\frac{m_n}{m_j}\right] \notag\\
&&\times \frac{\partial(\phi^{(1)})^2}{\partial\xi}.\label{vpny2-mkdv}
\end{eqnarray}
\subsection*{mKdv equation}
\begin{figure*}[ht]
\centering
\includegraphics[height=3.0in,width=6in]{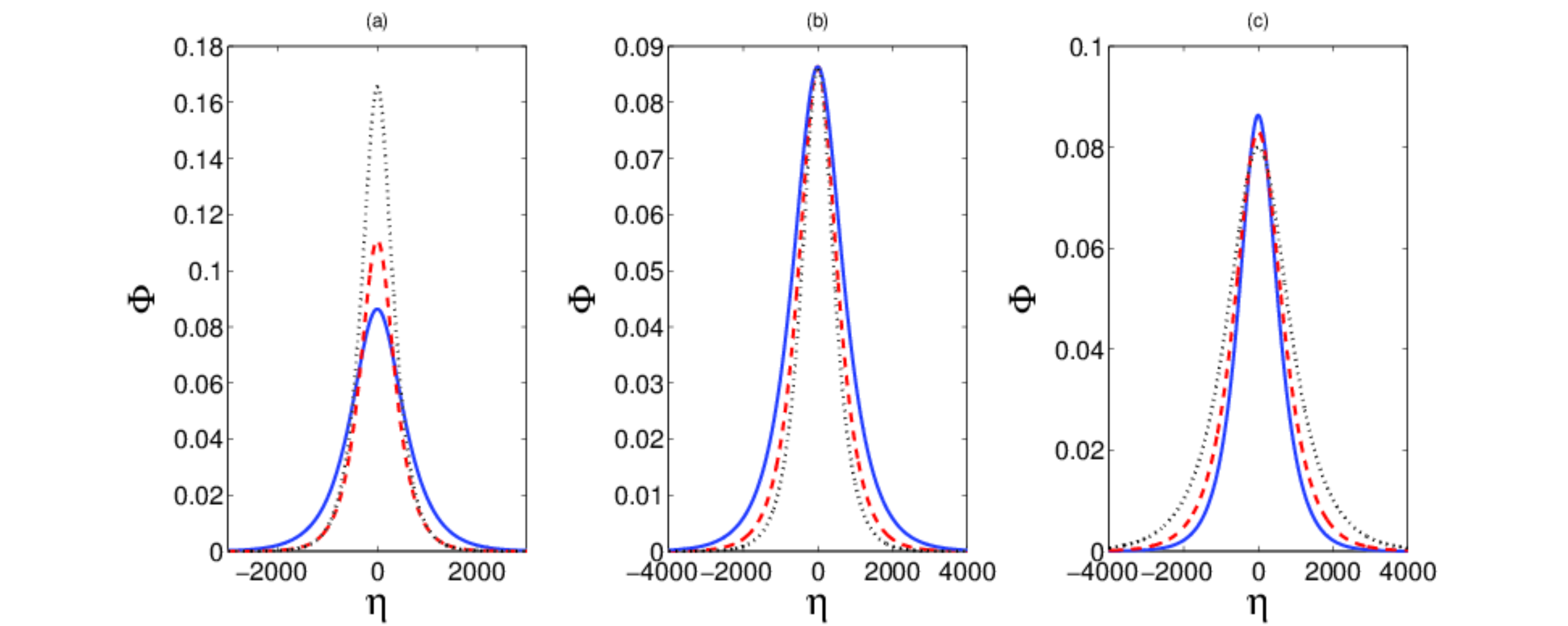}
\caption{Stationary soliton solution [Eq. \eqref{soliton-mkdv}] of the mKdV equation \eqref{mkdv} at $\tau=0$, and with $U_0=0.001$ is shown with the variations of (a) $\delta$: $\delta=0.9$ (solid line), $0.91$ (dashed line) and $0.92$ (dotted line); the other parameter values are $T=0.1$, $l_z=0.8$, $\omega_{cp}=200$ and $m=m_c-10$  (b) $\omega_{cp}$: $\omega_{cp}=150$ (solid line), $200$ (dashed line)  and $250$ (dotted line); the other parameter values are $\delta=0.9$, $T=0.1$, $l_z=0.8$ and $m=m_c-10$ (c) $T$: $T=0.1$ (solid line), $0.3$ (dashed line) and $0.5$ (dotted line); the other parameter values are $\delta=0.9$, $l_z=0.8$, $\omega_{cp}=200$ and $m=m_c-10$.     }
\label{fig:figure6}
\end{figure*}

We eliminate $v^{(3)}_{jz}$ and $n^{(3)}_j$ from Eqs. \eqref{cont-3mkdv}, \eqref{moment-3p-z-mkdv}, \eqref{moment-3n-z-mkdv} and \eqref{poisson-3-mkdv} and note that the coefficients of $\phi^{(3)}$ and $\partial\left(\phi^{(1)}\phi^{(2)}\right)/\partial\xi$  vanish, respectively, by means of Eq.\eqref{phase-velo} and the critical condition $A=0$. Thus, one obtains the following mKdV equation
\begin{equation}
\frac{\partial\Phi}{\partial\tau}+A'\Phi^2\frac{\partial\Phi}{\partial\xi}+B\frac{\partial^3\Phi}{\partial\xi^3}=0,\label{mkdv}
\end{equation}
where $\Phi\equiv\phi^{(1)}$. The  dispersion coefficient is the same as in Eq. \eqref{K-dV}, however,   the coefficient of nonlinearity which appears as a higher-order effect, is   given by
\begin{eqnarray}
A'=&&\frac{1}{10}\frac{l_z}{\mu^3}\left[\frac35 m(1+\mu T)\right]^{1/2} \frac{(1+\mu m)^{3/2}}{(m-T)^3}\notag \\
&&\times\left[\frac{45}{2}\left(\frac{1+\mu T}{1+\mu m}\right)\left(\mu^4m^2-1\right)+3\left(\mu^4mT-1\right)\right.\notag \\
&&\left.+\frac12\frac{\left(\mu^4T^2-1\right)(1+\mu m)}{1+\mu T}\right.\notag\\
&&\left.+\frac23\frac{\mu(1+\mu^3 T)(m-T)}{1+\mu T}\right].\label{nonlin-coeff-mkdv}
\end{eqnarray}
Equation \eqref{mkdv} describes the evolution of DIA waves in pair-ion plasmas in the critical condition     $m\sim m_c\gg1$, i.e., in plasmas containing much heavier negative ions than the positive ions. 
A stationary soliton solution of Eq. \eqref{mkdv} is obtained by applying a transformation $\eta=\xi-U_0\tau$ as
\begin{equation}
\Phi=\Psi~\text{sech}\left(\eta/W\right), \label{soliton-mkdv}
\end{equation}
where $\Psi=6U_0/A'$ is the amplitude and $W=\sqrt{B/U_0}$ is the width of the mKdV soliton with $B>0$. Inspecting on the coefficients $A'$ and $B$ we find that for typical values of the parameters with $m\sim m_c\gg1$, $\omega_{cp}\sim1$, $T<T_c<1$ and $\mu<1/9$, we have $A<1$ and $B\gg1$. Since soliton formation is due to a nice balance between the nonlinear and the dispersion coefficients, one should not expect $B\gg A$. Thus, in order to have appreciable values of $A$ and $B$, we consider $\omega_{cp}\gg1$.    Figure \ref{fig:figure6} shows the characteristic features of the mKdV soliton with the variations of (a) $\delta$, (b) $\omega_{cp}$ and (c) $T$. We find that as $\delta$ increases, the amplitude of the soliton increases, while the width decreases for $\delta\lesssim0.91$. As in the case of KdV solitons, the effect  of the external magnetic field, characterized by $\omega_{cp}$, is to decrease the soliton width only, since $A$ is independent of $\omega_{cp}$. We also find that as the temperature ratio $T$ increases, the width increases, while the amplitude of the soliton decreases. Similar to the KdV solitons, the  effects of $l_z$   is to decrease   (increase) the  amplitude (width) of the mKdV DIA solitons.   
\section{Discussion and conclusion}
We have investigated the characteristics of small-amplitude electrostatic perturbations propagating obliquely to the external magnetic field in an electron-free dusty pair-ion plasma. 
\begin{table}[ht]
\caption{Laboratory plasma parameters}  
\centering                     
\begin{tabular}{|c|c|}       
\hline                   
$m_p$ &$39m_{\text{proton}}=6.5\times10^{-23}$ g \\\hline 
$m_n$ & $146m_{\text{proton}}=2.4\times10^{-22}$ g\\\hline
$T_p$ & $0.2~$ev$=2321~$K\\\hline
 $T_n$ & $0.025~$ev$=290.12~$K\\\hline
$n_{n0}$ & $2\times10^9$ cm$^{-3}$\\\hline
 $n_{p0}$ & $1.3\times10^9$ cm$^{-3}$\\\hline
 $n_{d0}$ & $2\times10^6$ cm$^{-3}$ \\\hline
$\phi_s$ & $0.1$ v$=3.3\times10^{-4}$ statv\\\hline
 $R$ & $5.04~\mu$m$=5.04\times10^{-4}~$cm\\\hline
 $z_d$ & $\sim R\phi_s/e\sim350$ \\\hline
 $B_0$ & $0.3~$T$=3000~$G\\
[1ex]   
\hline 
\end{tabular}
\label{table:laboratory-plasma-parameters}    
\end{table}
\begin{table}[ht]
\caption{Space plasma parameters}
\centering                     
\begin{tabular}{|c|c|}       
\hline                   
$m_p$ &$28m_{\text{proton}}=4.7\times10^{-23}$ g \\\hline 
$m_n$ & $300m_{\text{proton}}=5.02\times10^{-22}$ g\\\hline
$T_p$ & $200~$K\\\hline
 $T_n$ & $200~$K\\\hline
$n_{n0}$ & $2\times10^6$ cm$^{-3}$\\\hline
 $n_{p0}$ & $10^6$ cm$^{-3}$\\\hline
 $z_dn_{d0}$ & $10^6$ cm$^{-3}$ \\\hline
$\phi_s$ & $0.7$ v$=2.3\times10^{-3}$ statv\\\hline
 $R$ & $0.6~$nm$=6\times10^{-8}~$cm\\\hline
  $B_0$ & $0.5~$G\\
[1ex]   
\hline 
\end{tabular}
\label{table:space-plasma-parameters}    
\end{table}

In the  linear regime (i.e., the very small-amplitude limit), we have Fourier analyzed the basic (linear) equations, and found that  low-frequency (in comparison with the  ion-cyclotron frequency) long-wavelength oblique slow and fast  modes can propagate as  dust ion-acoustic (DIA)  and dust ion-cyclotron (DIC)-like waves.   The properties of these waves are studied with the effects of (i) obliqueness of propagation angle with the magnetic field $(\theta)$,  (ii) charged dust impurity in the plasma $(\delta)$,  (iii) the static magnetic field $(\omega_{cp})$, (iv) the   adiabatic ion temperature ratio   $(T)$, and (v) the  ion mass ratio $(m)$ (See Fig. \ref{fig:figure1}). We show that the frequency gap between the two fast and slow modes    decrease as $k\rightarrow0$ by the effects of $\delta$ and $T$. The frequencies of   these modes are significantly altered under the influence of the external magnetic field and different masses of ions $(m\neq1)$. Furthermore, a simultaneous increase in the frequencies of both the modes is observed for   higher values of the temperature ratio $T$, and the magnetic field characterized by $\omega_{cp}$. Also, for a wave number higher than its critical value, an opposite trend of change in the frequencies of the modes is seen to occur by the effects of $\delta$ and the mass ratio $m$ $(>1)$. A similar analysis has also been carried out for the characteristics of DIA and DIC modes [Figs. \ref{fig:figure2} and \ref{fig:figure3}, which, in particular, appear for wave propagation   parallel and perpendicular to the magnetic field.

In the nonlinear theory, special emphasis is given to study the oblique propagation of electrostatic solitary waves. The transverse velocity perturbations of the ion fluids are assumed to be of higher-order effects than that for the parallel components. Such anisotropy is introduced because, the ion gyro-motion   is treated as a higher-order  effect than the motion along the magnetic field. Thus, we are interested  to consider the nonlinear propagation of DIA solitary waves (instead of DIC waves). A  standard reductive perturbation technique is used to  derive  a KdV equation which describes the evolution of electrostatic DIA perturbations.  We have  shown that the phase speed of the nonlinear wave in the moving frame of reference corresponds to that of the obliquely propagating  low-frequency linear DIA mode. Furthermore, it is found that the dispersion coefficient $B$ of the KdV equation is always positive, while the nonlinear coefficient $A$ can be  negative or positive  depending on whether the mass ratio $m$ is below or above its critical value $m_c$. The latter typically depends on the temperature ratio $T$ and the density ratio $\mu$.   It is found that for appreciable values of  $m_c~(>0)$, the conditions $T<T_c\equiv(1-9\mu)/8\mu^2<1$ and $\mu<1/9$ must be satisfied, and thus, one finds $m_c\gg1$. The latter corresponds to plasmas with much heavier negative ions than the positive ones $(m\gg1)$. However, for typical laboratory and space plasma parameters as in Tables \ref{table:laboratory-plasma-parameters} and \ref{table:space-plasma-parameters}, $m\ll m_c$ and  so the KdV equation admits only raraefactive DIA solitons. The formation of compressive solitary waves may be possible for $m\gg m_c$, however, those plasma regimes may not be relevant in laboratory and space environments. On the other hand,  for values of $m$ close to $m_c$, i.e., $m\sim m_c\gg1$, the KdV equation fails to describe the evolution of DIA solitons. In this case, we have derived a modified KdV equation which is shown to admit  only compressive DIA  solitons in pair-ion plasmas with much heavier negative ions than the positive ones.  It is to be noted that the critical mass $m_c$ for which the nonlinear coefficient $A$ changes its sign may not appear in a magnetized  pair-ion  plasma  without any charged dust impurity.  In this case solitary waves may exist  with  only the negative potential. Thus, charged dust grains in the background plasma play important roles for the existence of compressive or rarefactive solitons. 
\begin{table}[ht]
\caption{Frequency estimates for laboratory plasmas}
\centering                     
\begin{tabular}{|c|c|c|c|}       
\hline                   
Wave & $k\lambda_D/\lambda$ (cm)& $\omega_{\text{fast}}/2\pi$ (kHz)& $\omega_{\text{slow}}/2\pi$ (kHz)\\\hline 
&$0.01/4.7$&$48.5$& $10.5$\\ 
\raisebox{1.5ex}{Oblique}& $0.2/0.23$ & $366$ & $28.6$\\
& $0.7/0.07$ & $125.7$ & $28.6$\\\hline
&$0.01/4.7$&$1.4\times10^3$& $ 9.4$\\ 
\raisebox{1.5ex}{DIA}& $0.2/0.23$ & $1.5\times10^3 $ & $187.2$\\
& $0.7/0.07$ & $ 1.8\times10^3$ & $551.8$\\\hline
&$0.01/4.7$&$1.4\times10^3$& $ 68.5$\\
\raisebox{1.5ex}{DIC}& $0.2/0.23$ & $1.5\times10^3$ & $ 124.8$\\
& $0.7/0.07$ & $ 1.6\times10^3$ & $363.2$\\
[1ex]   
\hline 
\end{tabular}
\label{table:frequency-laboratory-plasma}    
\end{table}
\begin{table}[ht]
\caption{Frequency estimates for space plasmas}
\centering                     
\begin{tabular}{|c|c|c|c|}       
\hline                   
Wave & $k\lambda_D/\lambda$ (cm)& $\omega_{\text{fast}}/2\pi$ (kHz)& $\omega_{\text{slow}}/2\pi$ (kHz)\\\hline 
&$0.01/4.7$&$31.2$& $0.16$\\ 
\raisebox{1.5ex}{Oblique}& $0.2/0.23$ & $626.1$ & $0.16$\\
& $0.7/0.07$ & $2.2\times10^3$ & $0.16$\\\hline
&$0.01/4.7$&$1.9\times10^3$& $ 15.6$\\ 
\raisebox{1.5ex}{DIA}& $0.2/0.23$ & $2\times10^3 $ & $273$\\
& $0.7/0.07$ & $ 2.7\times10^3$ & $826.7$\\\hline
&$0.01/4.7$&$1.9\times10^3$& $ 7.8$\\
\raisebox{1.5ex}{DIC}& $0.2/0.23$ & $1.95\times10^3$ & $ 158.3$\\
& $0.7/0.07$ & $ 2.2\times10^3$ & $522.5$\\
[1ex]   
\hline 
\end{tabular}
\label{table:frequency-space-plasma}    
\end{table}

We, however,   mention that the present nonlinear theory of DIA waves is not valid for $T=m$, i.e, when the ratios assume equal values (in particular, for pair-ion plasmas with equal mass and temperature, i.e., $m=T=1$). The stationary soliton solutions of the KdV and mKdV equations are obtained, and their properties are analyzed numerically. It is found that the influence of the external magnetic field   is to make the soliton narrower in reducing its width   without any change in the amplitude. Similar to the effects of charged dusts, the ion-temperature  ratio has also the effect to decrease the soliton amplitude along with a slight increase in its width. The effect of the mass ratio on the soliton profile is almost similar to that of $T$, however, a significant change in the width is seen to occur. 

 Our theoretical results may be used for   experimental verification. For example, we may consider a laboratory  plasma \cite{pair-ion-charging-experiment,pair-ion-instability}  [See Table \ref{table:laboratory-plasma-parameters}] in which the light positive ions are singly ionized potassium $K^{+}$ and the heavy negative ions are $SF_6^{-}$. Thus, the negative to positive ion mass ratio is $m\approx146/39\approx3.74$, the negative to positive ion temperature ratio is $T\approx1/8=0.125$, the negative ion number density is $n_{n0}\sim2\times10^9$ cm$^{-3}$, so that $\mu<1$. Also, according to Kim and Merlino \cite{pair-ion-charging-experiment}, when $n_{p0}\sim500n_{e0}$, where $n_{e0}$ is the electron number density, dusts can be positively charged to a surface potential $\phi_s\sim0.1$ V. For a grain of radius $R=5~\mu$m, the charge state $z_d\sim R\phi_s/e\sim350$. Thus, if $n_{n0}\sim10^3n_{e0}$ and $n_{p0}\sim650n_{e0}$, so that $\mu=0.7$, then   the charge neutrality condition $n_{p0}+z_dn_{d0}=n_{n0}$ requires $n_{d0}\sim2\times10^6$ cm$^{-3}$. Suppose that the plasma is immersed in a magnetic field of strength $B_0\sim0.3~$T, then for $T_p=0.2~$ev and $T_n=T_p/8$,  we have, $T=0.12$, $v_{tp}/c_s=1.9$, $v_{tn}/c_s=0.35$, $\omega_{cp}/\omega_{pn}=0.15$, $\omega_{cn}/\omega_{pn}=0.04$, $\lambda_D=0.0074~$cm, $\omega_{pn}=4.9\times10^6~$s$^{-1}$, $c_s=3.6\times10^5~$cm$/$s.  Thus, one can estimate frequencies of different kind of wave modes at different wavelengths $(\lambda=2\pi/k)$ as in Table \ref{table:frequency-laboratory-plasma}. Alternatively, one can also have similar estimates from the graphical representations of the wave modes as in Figs. \ref{fig:figure1} to \ref{fig:figure3}. Furthermore, considering the   plasma parameters as in Table \ref{table:laboratory-plasma-parameters}, one can also calculate the nonlinear wave speed given by Eq. \eqref{phase-velo}, the nonlinear and dispersion coefficients $A$ and $B$, as well as the    amplitudes and width of the solitons. 

On the other hand, Rapp et al. \cite{in-situ-Rapp} suggested that the presence of sufficiently heavy and numerous negative ions (i.e., $m_n>300$ amu and $n_{n0}\gtrsim50n_{e0}$) can explain their observations of positively charged dusts in the Earth's mesosphere between 80 and 90 km. Thus, for space plasma environments [See Table \ref{table:space-plasma-parameters}], e.g.,  a dusty region at an altitude of about 95 km, we can have $T_p\sim T_n\sim200$ K, $m_n/m_p=300/28\approx10.7$, $n_{n0}\sim2\times10^6$ cm$^{-3}$, $n_{p0}\sim10^6$ cm$^{-3}$, $z_dn_{d0}\sim10^6$ cm$^{-3}$. Thus, for a magnetic field strength $B_0\sim0.5$ G, we calculate 
$T=1$, $v_{tp}/c_s=3.3$, $v_{tn}/c_s=1$, $\omega_{cp}/\omega_{pn}=0.002$, $\omega_{cn}/\omega_{pn}=1.5\times10^{-4}$, $\lambda_D=0.07~$cm, $\omega_{pn}=1.1\times10^5~$s$^{-1}$, $c_s=7.4\times10^3~$cm$/$s. The frequency estimations of different modes at different wavelength are shown in Table \ref{table:frequency-space-plasma}. As above we can also estimate the nonlinear wave speed and the soliton characteristics for space plasma parameters as in Table \ref{table:frequency-space-plasma}.

To conclude, we have studied the effects of immobile positively charged dusts, the obliqueness of propagation, the external magnetic field, the pressure gradient forces and different masses of ions, and their significance for the excitation of oblique DIA and DIC-like waves    as well as the nonlinear evolution of  KdV and mKdV DIA solitons in magnetized  dusty pair-ion plasmas including the case with much heavier negative ions. The theoretical results may be useful for the observation of different kind of wave modes including  DIA and DIC waves together with their coupling as well as the nonlinear evolution of DIA solitary waves and solitons   in laboratory and space plasmas, e.g.,   in the Earth's mesosphere, a magnetized dusty negative-ion plasma region at an altitude of about $95$ km \cite{in-situ-Rapp}.  The results may also be applied to other plasma environments comprising magnetized multi-ions   with positively charged dusts, and the  parameters satisfy $m\gg1$ and $T,~\mu<1$.

 It is to be noted that the inclusion of the effects of ion-dust collisions,  ion-drag forces due to positive and negative ions on the charged dusts, as well as the ion-kinematic viscosities in the present model could be another problem of interest, but beyond the scope of the present investigation. Furthermore, the electrostatic disturbances in magnetized pair-ion plasmas with the dynamics and charging of  massive dusts could be an  another interesting piece of work. However,   the processes of charging of dust particles, especially  in space plasmas, e.g., in mesospheric nanoparticles are much more complicated that hitherto assumed \cite{in-situ-Rapp}. Thus, more in situ, laboratory, and theoretical investigations are needed to study the size distribution and charging properties of mesospheric nanoparticles,  and  their significance for the propagation of DIA waves and other nonlinear phenomena.

\section*{Acknowledgement}
{A. B. is thankful to University Grants Commission (UGC), Govt. of India, for Rajib Gandhi National Fellowship with Ref. No. F1-17.1/2012-13/RGNF-2012-13-SC-WES-17295/(SA-III/Website).  This research was partially supported by   the SAP-DRS (Phase-II), UGC, New Delhi, through sanction letter No. F.510/4/DRS/2009 (SAP-I) dated 13 Oct., 2009, and by the Visva-Bharati University, Santiniketan-731 235, through Memo No.  REG/Notice/156 dated January 7, 2014.}

\end{document}